\documentclass[jcp, floatfix, nobibnotes, reprint, superscriptaddress]{revtex4-1}
\usepackage{docs}%
\usepackage{siunitx}
\usepackage{amsmath}
\usepackage{booktabs}
\DeclareSIUnit[number-unit-product = {\,}]
\cal{cal}
\DeclareSIUnit\kcal{\kilo\cal}
\DeclareSIUnit[number-unit-product = {\,}]
\Btu{Btu}
\DeclareSIUnit[number-unit-product = {\,}]
\Fahr{\degree F}
\DeclareSIUnit[number-unit-product = {\,}]
\usepackage{bm}
\usepackage[colorlinks=false,linkcolor=black]{hyperref}%
\usepackage{graphicx}
\usepackage[utf8]{inputenc}
\usepackage{tabularx}

\usepackage{dcolumn}
\usepackage{epstopdf}
\usepackage{afterpage}
\usepackage{setspace}
\usepackage{multirow}
\usepackage{dsfont}
\usepackage{sidecap}
\usepackage{epstopdf}
\usepackage{braket}
\usepackage{csquotes}
\usepackage{cleveref}

\usepackage{siunitx} 
\usepackage{color, colortbl}

\definecolor{Gray}{gray}{0.9}

\AppendGraphicsExtensions{.gif}

\DeclareMathAlphabet\mathbfcal{OMS}{cmsy}{b}{n}

\setcitestyle{super}

\begin{document}

\setstretch{1.0}

\title{Adaptive hybrid density functionals} 

\author{Danish Khan}
\affiliation{Chemical Physics Theory Group, Department of Chemistry, University of Toronto,
St. George Campus, Toronto, ON, Canada}
\affiliation{Vector Institute for Artificial Intelligence, Toronto, ON, M5S 1M1, Canada}

\author{Alastair J. A. Price}
\affiliation{Chemical Physics Theory Group, Department of Chemistry, University of Toronto,
St. George Campus, Toronto, ON, Canada}
\affiliation{Acceleration Consortium, University of Toronto, Toronto, ON, Canada}

\author{Maximillian L. Ach}
\affiliation{Department of Physics, University of Toronto, St. George Campus, Toronto, ON, Canada}
\affiliation{Department of Physics, Ludwig-Maximilians-Universität München (LMU), Munich, Germany}
\author{O. Anatole von Lilienfeld}
\email{anatole.vonlilienfeld@utoronto.ca}
\affiliation{Chemical Physics Theory Group, Department of Chemistry, University of Toronto,
St. George Campus, Toronto, ON, Canada}
\affiliation{Vector Institute for Artificial Intelligence, Toronto, ON, M5S 1M1, Canada}
\affiliation{Acceleration Consortium, University of Toronto, Toronto, ON, Canada}
\affiliation{Department of Physics, University of Toronto, St. George Campus, Toronto, ON, Canada}
\affiliation{Department of Materials Science and Engineering, University of Toronto,
St. George Campus, Toronto, ON, Canada}
\affiliation{Machine Learning Group, Technische Universit\"at Berlin and Institute for the Foundations of Learning and Data, 10587 Berlin, Germany}
\affiliation{Berlin Institute for the Foundations of Learning and Data, 10587 Berlin, Germany}

\vspace{1mm}
\begin{abstract}
\section*{Abstract}
\vspace{-2mm}
Exact exchange contributions are known to crucially affect electronic states, which in turn govern covalent bond formation and breaking in chemical species.
Empirically averaging the exact exchange admixture over compositional degrees of freedom, hybrid density functional approximations  have been widely successful, yet have fallen short to reach 
high level quantum chemistry accuracy, primarily due to delocalization errors. 
We propose to `adaptify` hybrid functionals by generating optimal admixture ratios of exact exchange on the fly, i.e.~specifically for any chemical compound, using extremely data efficient quantum machine learning models that carry negligible overhead. 
The adaptive Perdew-Burke-Ernzerhof based hybrid density functional (aPBE0) is
shown to yield atomization energies with sufficient accuracy to 
effectively cure the infamous spin gap problem in open shell systems, such as carbenes. 
aPBE0 further improves energetics, electron densities, and HOMO-LUMO gaps in organic molecules
drawn from the QM9 and QM7b data set. 
Obtained with aPBE0 in a large basis, we present a revision of the entire QM9 data set (revQM9) containing more accurate quantum properties with on average,
stronger covalent binding, larger band-gaps, more localized electron densities, and larger dipole-moments. 
While aPBE0 is applicable in the equilibrium regime, outstanding limitations include
covalent bond dissociation when going beyond the Coulson-Fisher point. 
\end{abstract}
\maketitle

\section*{Introduction}
Accelerating the inverse design of novel materials and chemicals
is a pressing concern for a more sustainable future.\cite{back2024accelerated} 
Quantum mechanics underpins our ability to predict electronic, optical, and thermal properties
with high fidelity, essential for designing materials with specific
functionalities.\cite{ceder1998predicting,franceschetti1999inverse,zunger2018inverse,smallmolinverse}
While universal in principle, numerically solving the 
many-body electronic Schr\"odinger equation has remained 
an outstanding challenge for all but the simplest systems
due to its immense computational complexity. 
Even the most recent and promising machine learning based
quantum models, \cite{carleo2017solving,ferminet,paulinet}
still rely heavily on considerable training data needs,  
require the optimization of millions of regression weights, 
and struggle to efficiently extrapolate across chemical compound space. 
By contrast, density functional theory (DFT) is a formally exact quantum method
but unfortunately the exact density functional is only known to exist, 
with no prescription to be found.~\cite{mattsson2002pursuit} 
Nevertheless, DFT enables the definition and parametrization of
density functional approximations (DFAs) which 
stand out in terms of offering the most viable trade-off between predictive power and
computational cost--in addition to their high degree of
reproducibility--\cite{lejaeghere2016reproducibility}, 
which effectively made DFT the principal workhorse for 
atomistic simulations in biology, chemistry, and materials. 
These qualities of DFT have also proven crucial for the rapidly growing
influx of AI in chemistry and materials simulation, enabling the tremendous
acceleration of routine computational materials and molecular discovery
campaigns we are witnessing today.\cite{dftai} 

Many approximations to the exact functional have
been made over the years~\cite{strayingperdew,burke2012JCP,becke50years}
and while currently the most reasonable tool for an impressive
array of compound classes and properties, common DFAs still 
struggle to reach the
accuracy of other post-HF methods.\cite{cohen2008insights,strayingperdew,wires}
A large number of such shortcomings 
stem from what is known as self-interaction error, which arises
from the Hartree energy term that includes the interaction of the electron
with itself.~\cite{becke50years} 
Self-interaction, or delocalization errors in general, manifest in a number
of well known ways, from the poor performance on band gap
energies,\cite{sham1985density,perdew1985density,cohen2008fractional} to the
prediction of spin gaps, critical to
the understanding of reactivity and magnetism, which can be both quantitatively and qualitatively incorrect.\cite{hostavs2023important} 
Rooted rigorously in adiabatic connection,~\cite{burke1997adiabatic} 
some of these short-comings can be mitigated by use of  
hybrid functionals that blend in 
HF exchange (also referred to as exact exchange) with the DFA exchange energy, 
\begin{equation} E_\mathrm{X}^\mathrm{hybrid} =
aE_\mathrm{X}^\mathrm{HF} +
(1-a)E_\mathrm{X}^\mathrm{DFA},
\label{hybrid} 
\end{equation} \noindent where $a$ is the mixing parameter, 
varying between 0 and 1. 
Global hybrids, originally parameterized for general thermochemistry, 
contain a fixed amount of exact exchange, typically around 20-25\%. 
Although legacy hybrid functionals such as B3LYP\cite{b3} or PBE0\cite{pbe0,PBE01} represented
significant advancements for energetics and electron densities,
they are not a panacea: 
More recently proposed DFAs were noted to sacrifice 
the accuracy of calculated electron densities for improved
energy predictions.~\cite{strayingperdew,brorsen2017accuracy} 
Common hybrid based DFA parameterizations attempt to find the average
optimum of a single universal admixture ratio to all chemistries and properties. 
However, Langreth and Perdew observed already in 1975~\cite{langreth1975exchange} that 
non-uniform systems can be modelled extremely well by tuning the exact exchange. 
The system dependency of optimal exchange admixture ratios has also been
noted by Johnson\cite{wires}. 
For example, even the most recent functionals do not reach 
chemical accuracy when it comes to reaction energies, barrier heights, 
thermochemistry, or systems with large self-interaction errors.~\cite{poison}
The same authors also point out that, at the expense of being specific
to each system, optimal exchange admixture ratios can drastically reduce the error. 
Admixture ratios are also a persistent issue for molecular crystals, where increasing amounts of HF exchange can be required to recover experimentally observed structures as the lowest energy on the crystal energy landscape.\cite{price2023xdm,price2023accurate}
Large delocalization errors were also identified to be the cause behind the first reported case in which DFA calculations predicted the incorrect structure for organic acid-base co-crystals which was fixed by inclusion of close to $\sim$42\% exact exchange.\cite{cocrystal_erin2018}
Further cases are highlighted by the same authors \cite{price2023accurate} showing that delocalization error is an issue for a number of chemically relevant systems, such as charge-transfer complexes,\cite{ruiz1996charge, sini2011evaluating, steinmann2012interaction} charge transfer excitation energies,\cite{cai2002failure, tozer2003relationship, dreuw2003long} halogen bonded complexes,\cite{otero2014halogen} barrier heights of radicals,\cite{lynch2001well, janesko2008hartree} and molecular crystals where up to $\sim 50 \%$ exact exchange mixing can be required to recover accurate energies and geometries.
Within another large scale study, the importance of exact exchange admixture for
correctly reproducing the correct high-spin low-spin ordering in first and second
row transition metal complexes was noted.\cite{nandy2020large}
But also the quality of hybrid DFA based molecular adsorption energies on transition metal oxide surfaces was shown to be strongly admixture dependent.\cite{zhao2019stable}
Santra and Martin highlighted  substantially improved performance of several DFAs by changing the admixture ratios when dealing with different chemistries from the GMTKN55 benchmark set.~\cite{santra2021types,gmtkn55}
Pasquarello and co-workers demonstrated dramatic improvements of band-gap predictions in insulators and semiconductors by tuning the amount of exact exchange.~\cite{yang2023range}
Also, Mewes, Grimme, and co-workers recently high-lighted the potential of tuning range-separated hybrid functionals.~\cite{friede2023optimally}
Last but not least, varying amounts of exact exchange in molybdenum carbide
systems resulted in direct reordering of spin-energies.\cite{hostavs2023important}

Based on these observations, we hypothesize the optimal HF/exact exchange admixture ratio, $a_\mathrm{opt}$, to be a system-specific scalar label  which approaches zero DFA error with respect to a trust-worthy high level electronic structure treatment (see Fig 1) and with sufficient smoothness characteristics to be modelled by machine learning. 
As long as this hypothesis holds, we note that system-specific optimal admixture ratios will  not violate energy conservation as they simply reproduce any energy conserving reference level of theory. 
Below, we present overwhelming numerical evidence in support of our 
hypothesis, and demonstrate that we can efficiently regress and estimate $a_\mathrm{opt}$ throughout chemical compound space using suitable machine learning models~\cite{CM_and_qm7,huang2021abinitio} (Fig 1).
Based on the recently introduced compact many body distribution functional (MBDF) representation,~\cite{mbdf} our KRR based quantum machine learning ansatz
is  data efficient and light weight in mapping directly from the 4$N$-6 internal degrees of freedom that define the electronic system to one scalar parameter in the Kohn-Sham Hamiltonian that only varies mildly in system changes.
This differs fundamentally from previously published ML based efforts aimed at improving DFAs which are significantly more challenging since they map the full electron density information, and vary during the self-consistent field cycles, to energy expectation values which are sensitive and strongly varying labels. 
In particular, in 2017  a machine learning model of the Hohenberg-Kohn map from external potential to electron density to energy was proposed.~\cite{Brockherde2017}
In 2020, neural network models were shown to successfully map electron density based descriptors to energies and forces,~\cite{dick2020machine,nagai2020completing} while 
the more recent DM21 functional by Kirkpatrick \textit{et al}~\cite{dm21} accounts 
for crucial additional constraints on fractional electron number and spin by 
learning the map from the spin indexed electron charge, gradient, kinetic energy, and two range separated HF densities to predict local enhancement factors in the exchange-correlation functional.
Concurrently, Margraf and Reuters presented a pure kernel ridge regression functional model predicting electron correlation using the HF density as an input,~\cite{Margraf2021} similar to other orbital based representations.~\cite{welborn2018transferability,karandashev2022orbital} 
The $\Delta$-ML~\cite{Ramakrishnan2015} based functional by Bogojeski \textit{et al}.~\cite{Bogojeski2020} also maps electron density to an energy (enabling force predictions). 
While more light weight, transferable and scalable,
by virtue of its embedding within standard KS-DFT
SCFs, our adaptive hybrid functional generates at least the same wealth of information 
(orbitals, eigenvalues, densities, energies) as all the other ML based functionals mentioned. 
Furthermore, it can conveniently be integrated in widely spread DFT codes which would facilitate its adaptation by a large international community of users, estimated to publish at least 30'000 papers each year.~\cite{pribram2015dft}

\begin{figure*}[htb]
          \centering           
          \includegraphics[width=\linewidth]{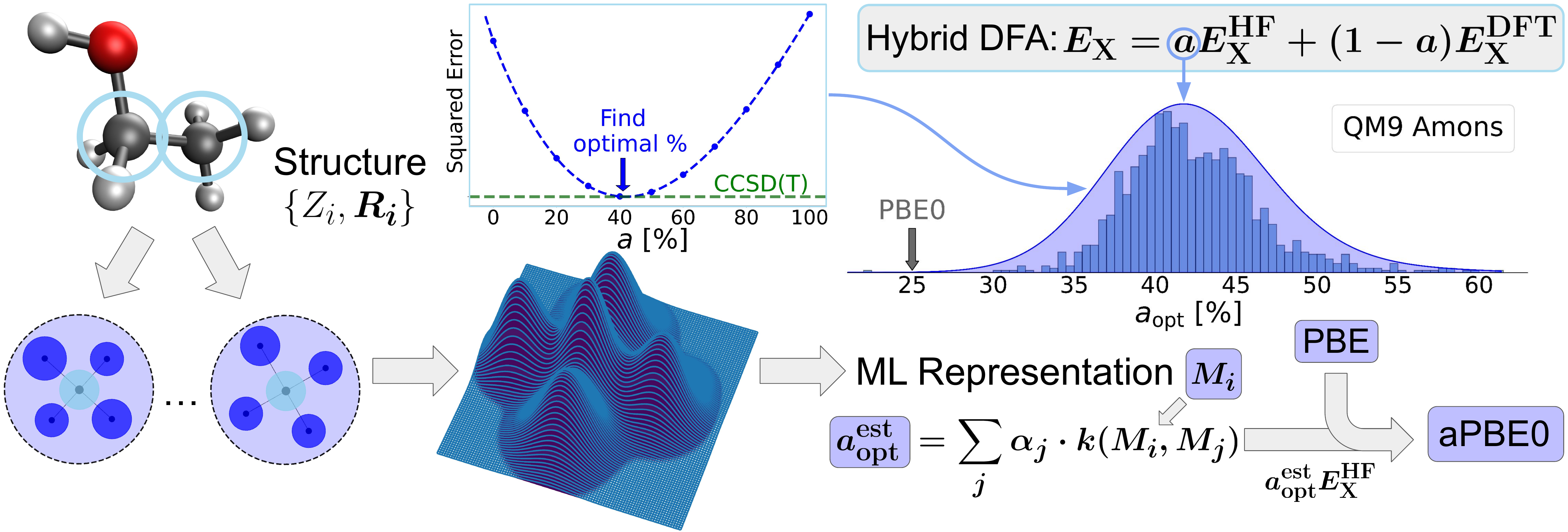}
          \caption{Workflow for the aPBE0 functional. Top left: Optimal HF admixture ratios, $a_\mathrm{opt}$, are obtained for a training set of molecules by minimizing the squared error (SE) of the atomization energy with respect to a high level reference, such as CCSD(T).
          Bottom left: A compound's nuclear charges and coordinates (defining the external potential in the electronic Hamiltonian) map to a suitable representation which, thanks to the kernel trick, enables linear regression using non-linear feature similarity measures.
          Right: for any out of sample molecule ($M_i$), the optimal HF exchange admixture ratio, $a_\mathrm{opt}$, can then be estimated by the ML model, leading to improved predictions for several observables at the same time. By contrast, previous applications of supervised learning would require individual respective training labels and models.    
          }
     \label{fig:workflow}
 \end{figure*}

In the following, we demonstrate that the idea of adaptive HF admixtures is effective for various properties. 
To this end, and in line with the principle of parsimony, we have chosen to adapt the PBE0~\cite{pbe0,PBE01} functional since it is based on the non-empirical GGA PBE~\cite{pbe}
and since there is only a single parameter blending in the HF exchange. 
We rely on quantum machine learning models to predict this single parameter on the fly, i.e. before launching the self-consistent field cycle, and using exclusively nuclear charges and atomic coordinates as an input~\cite{dftai}.  
The resulting ``adaptive'' PBE0 functional (aPBE0) improves upon all commonly used DFAs for obtaining qualitatively and quantitatively correct singlet-triplet spin gaps in carbenes.
For closed shell molecules, aPBE0 based forces hardly change, and we even find simultaneous improvements in  properties other than just atomization energies, namely in electron densities and frontier orbital gaps, indicating that the method does not sacrifice accuracy of one observable in favour of the other.\cite{strayingperdew}
After training on atomization energies calculated with Coupled Cluster with Singles Doubles perturbative Triples excitations (CCSD(T)), we have used aPBE0 to revise the entire QM9 dataset,\cite{QM9}, a popular molecular benchmark data set of quantum properties for over 130'000 small to medium sized biochemically relevant organic molecules, originally generated within the exhaustive molecular graph enumeration efforts by Reymond and co-workers,~\cite{GDB17}.
Finally, we explore the extension to adaptive electron correlation in the context of going beyond the Coulson-Fisher point within the bond dissociation energy profile. 

\section*{Spin gaps}

\begin{figure*}[!htb]
    \centering
    \includegraphics[width=\linewidth]{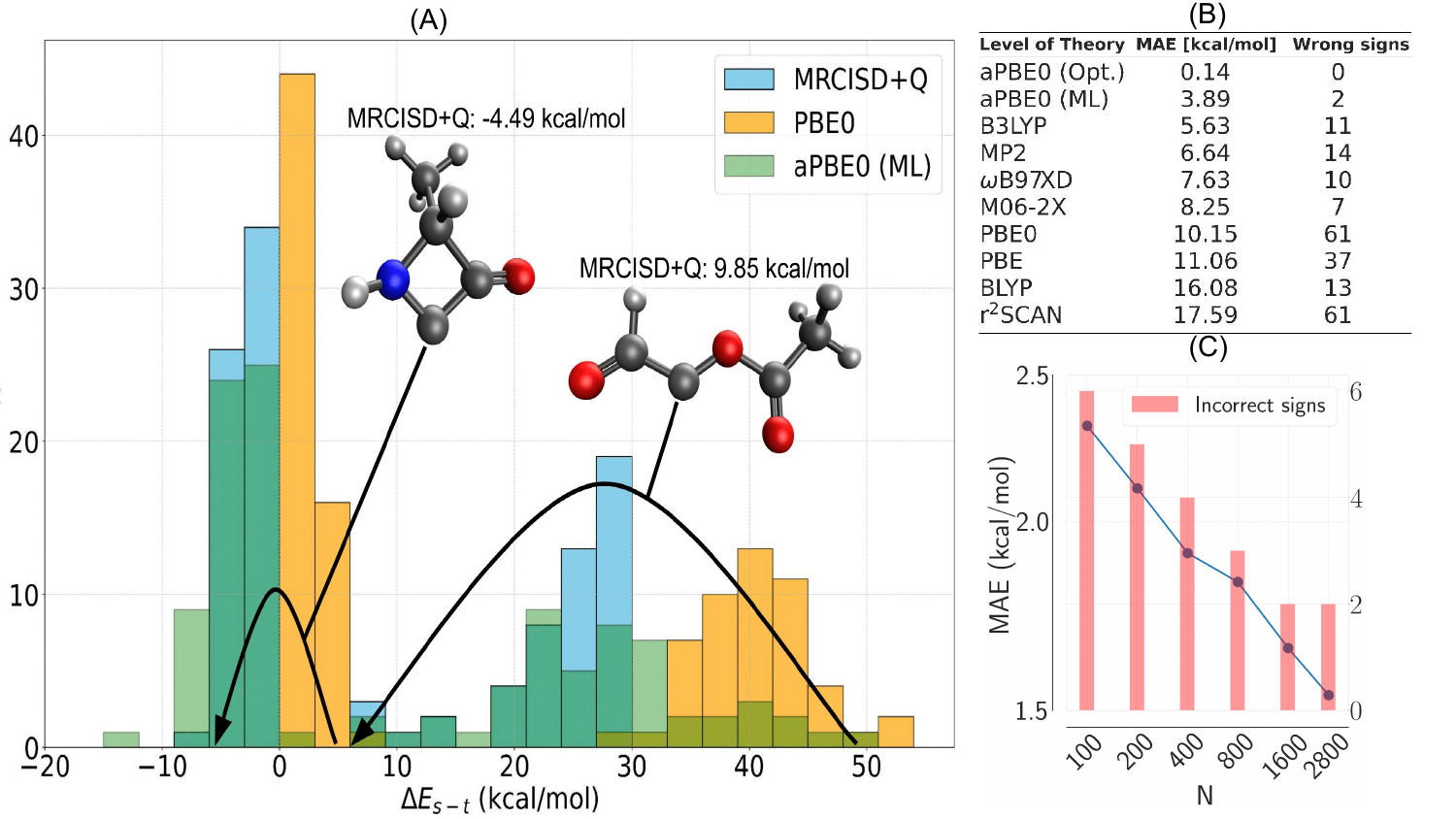}
    \caption{Estimates of vertical singlet-triplet spin gaps in organic radicals (carbenes).       (A) Distribution of spin gaps according to MRCISD+Q, PBE0 and aPBE0 level of theories for the 110 most challenging out-of-sample carbenes from the QMspin dataset\cite{qmspin}.
    The two most extreme carbenes (maximal deviation, wrong sign) are shown as an inset, with
    arrows indicating the shift from PBE0 to aPBE0.
    (B) Average quantitative (MAE) and qualitative (number of wrong signs) spin gap errors compared to MRCISD+Q values for the entire test set of 110 carbenes using various level of theories.
    aPBE0 (Opt.) refers to the PBE0 functional employing the optimal HF exchange fraction for all molecules found via optimization, while aPBE0 (ML) uses the HF exchange fraction predicted on the fly by our ML model (previously trained on $a_{\rm opt}$ values for 2800 other carbenes).
    (C) aPBE0 (ML) performance curve: Averaged gap prediction errors and number of incorrect signs (bar plot) decay with training set size $N$ for a validation set of 50 random out-of-sample carbenes (not part of training). 
    All DFT numbers are based on restricted open shell calculations, 
    for MP2 the unrestricted method was used.
    } 
    \label{fig:sping_perf}
\end{figure*}
To probe the potential of adaptive hybrids, we  consider one of the most challenging
quantum properties for popular DFAs: The spin gap energy in open shell systems,  commonly encountered in magnetic materials such as transition metal complexes (high-spin/low-spin), or in organic radicals.~\cite{hostavs2023important} 
More specifically, we rely on singlet-triplet gap information calculated at 
multireference MRCISD+Q level of theory and reported for thousands of carbenes, 
molecular organic radicals with two unshared electrons on a carbon atom,
as reported within the QMspin data set.~\cite{qmspin} 
Fig.~2A shows singlet-triplet spin gap distributions on a select out-of-sample test set (never used for training) of 110 most challenging carbenes: 
PBE0 predicts either very  {\em large quantitative} deviations, 
or even the wrong sign ({\em qualitatively} incorrect).
Two extreme cases, seemingly innocent radicals, are shown  in the inset of Fig.~2A,
with PBE0 giving a vertical spin gap value
of 4.9 kcal/mol, instead of -4.49 kcal/mol, for the one, and 
49.8 kcal/mol for the other, instead of the 9.85 kcal/mol reference value. 
After training separate models of $a_{\rm opt}$ 
for respective singlet and triplet states on up to $\sim$2'800 other carbenes, stored first in the QMspin data base, 
the difference between the aPBE0 based predictions of the singlet and triplet state energy
improves systematically and dramatically.
For the largest training set size, the estimated gap of singlet and triplet
atomization energies of the two molecules in questions correspond to -4.4 and 7.1 kcal/mol, respectively. 
But also the overall gap value distribution for the 110 challenging carbenes shifts markedly in the right direction when using aPBE0 after training on the largest training set, reducing the average error and the number of incorrect signs. 

Since delocalization errors are relevant for poor predictions of spin gaps,~\cite{hostavs2023important} we also present averaged errors for these 110 radicals
for the most commonly used DFAs from different rungs of Perdew's Ladder for comparison (Fig 2B).
Clearly, inclusion of HF exchange improves the performance, as  hybrid functionals generally 
outperform the most commonly used GGAs/meta-GGAs, and closely match or even outperform post-Hartree Fock wavefunction based 2nd order perturbation theory (MP2).
The average aPBE0 error of 3.9 kcal/mol (after training on $\sim$2800 other radicals) is substantially lower than for PBE0 (10.15 kcal/mol), and significantly 
better than the best functional B3LYP (5.6 kcal/mol), or even MP2 (6.6 kcal/mol). 
aPBE0 also produces the least number of wrongly predicted signs (2), 5 less than the best DFA (M06-2X). 
Closer inspection of the two carbenes for which the gap is still predicted with the wrong sign reveals that the gap is very close to zero, rendering oscillations of the sign less problematic.  
Furthermore, we note that the performance curve in Fig 2C does not look converged, suggesting that aPBE0 could be further improved through the use of even larger training set sizes. In particular, if sufficient training molecules were available,
our  learning curves (see Supplementary Materials) amount to numerical evidence that the optimal admixture ratio can be predicted with systematically improving statistical uncertainty, ~\cite{vapnik1994learningcurves},
resulting in systematically improving spin gap estimates.
Using optimal HF exchange values, the corresponding aPBE0 (Opt.) functional provides
nearly exact predictions in all cases with a vanishingly small 
average spin gap error, and all signs predicted correctly (Fig 2B).

\section*{Energies, densities, HOMO-LUMO gaps}
\begin{figure*}
    \centering
    \includegraphics[width=\linewidth]{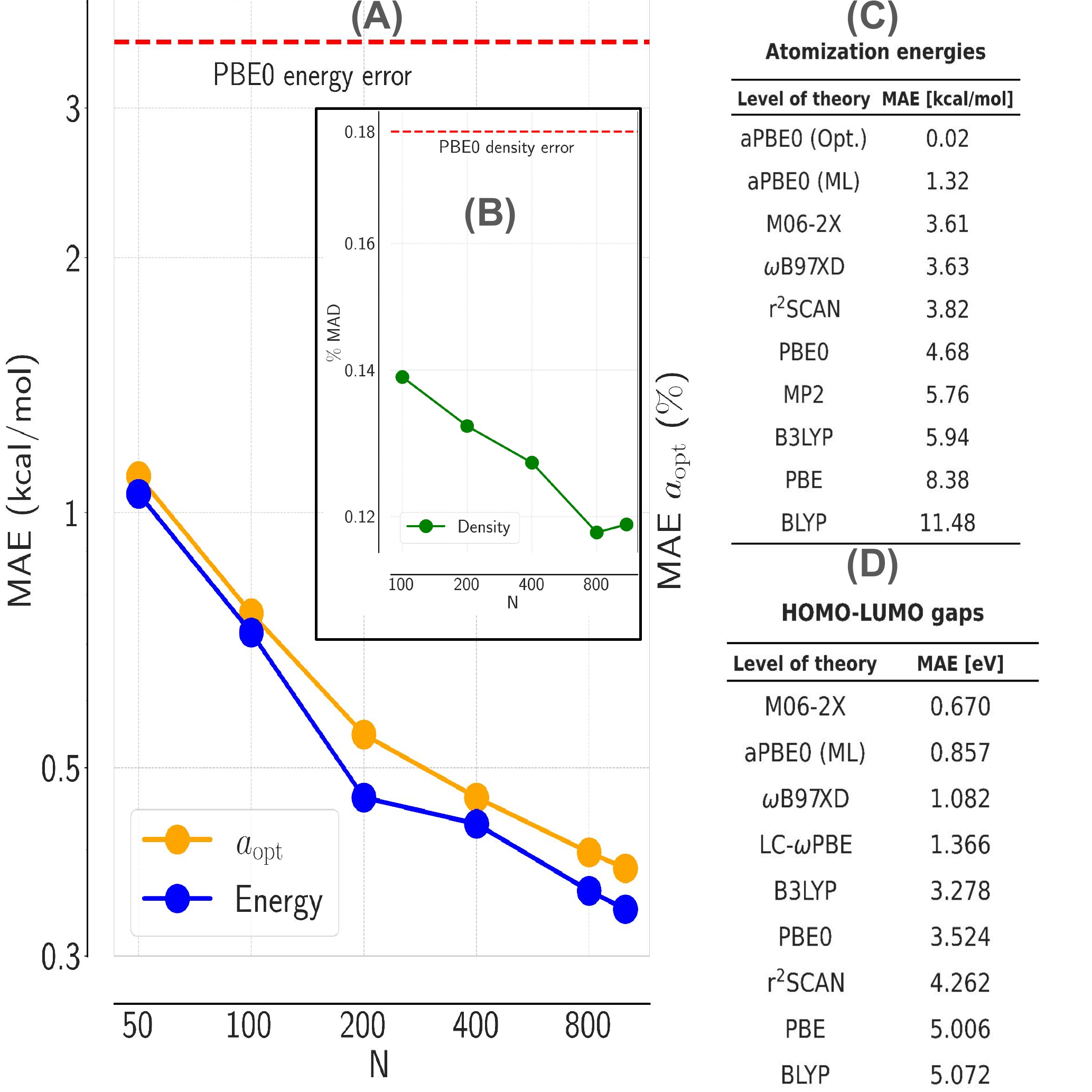}
    \caption{Transferability tests of aPBE0 as a foundation model treating various properties and organic compound distributions. All HF admixture ratios predicted with ML after training on optimal values for up to 1169 amon~\cite{amons} based small organic molecular fragments in Ref.~\cite{bing_DMC} containing not more than 5 heavy atoms.
        (A) Learning and performance curves for 100 out-of-sample amons showing respectively prediction errors in optimal HF admixture ratios $a_{\mathrm{opt}}$ and  
         atomization energy compared to CCSD(T)/cc-pVTZ values.
        (B) Deviations ($\sqrt{\int dr (\rho_{\rm aPBE0}-\rho_{\rm CCSD})^2/\int dr \rho_{\rm CCSD}^2}$) from  CCSD electron densities for a test set of 50 larger QM9 molecules with 9 heavy atoms each published in Ref.~\cite{bing_DMC}. 
    (C) Sorted atomization energy errors compared to CCSD(T)/cc-pVTZ values for the test set of 50 QM9 molecules using  different DFAs and MP2. 
    (D) HOMO-LUMO gap errors compared to GW\cite{gw} computed values for 100 organic molecules drawn at random QM7b dataset~\cite{qm7b}.
    }
    \label{fig:atm_perf}
\end{figure*}

To further assess if aPBE0 is transferable, 
or could even be considered a foundational model of ML in DFT,
we have studied its predictive performance for two other data-sets and observables, 
namely atomization energies and electron densities of 50 QM9 molecules from Ref.~\cite{bing_DMC}, and HOMO-LUMO gaps of the first 100 organic molecules recorded in 
the QM7b dataset~\cite{qm7b}.
In particular, these predictions were made using the same aPBE0 model, and after training on  $a_{\mathrm{opt}}$ values optimized using previously published CCSD(T) atomization energies of the 1169 amon\cite{amons} based small organic molecular fragments containing not more than 5 heavy atoms.~\cite{bing_DMC} (Figures 1, 2 in SI).
The corresponding distribution of $a_{\rm opt}$ values is centered at $\sim$42\% and is on display in Fig. 1.
The resulting learning curve in figure 3A indicates that the ML model of $a_{\rm opt}$ for out-of-sample amons systematically improves with amon training set size, reaching close to 0.35\% prediction error after training on 1'000 instances. 
Correspondingly improving performance errors of amon atomization energies on hold-out amons are also shown in Fig 3A, indicating that chemical accuracy is already reached after training $a$ on just 50 instances.

After training the most accurate $a$ model on all the 1'169 amons we then performed aPBE0 calculations of the 50 QM9 molecules with 9 heavy atoms used in Ref.~\cite{bing_DMC}. 
In comparison to PBE0, the predicted averaged absolute error (with respect to CCSD(T)) of the aPBE0 based atomization energy is much reduced, down from 4.68 kcal/mol for PBE0, to 1.32 kcal/mol (Fig 3C). 
Just as in the case of the spin gaps, a comparison to other common DFAs and MP2 indicates that
aPBE0 is the most accurate, significantly outperforming the next best functionals $\omega$B97XD (3.63 kcal/mol) as well as M06-2X (3.61 kcal/mol). 
Again, we include in Fig 3C aPBE0 results obtained for the truly optimal HF exchange ratio (aPBE0 (Opt.)) which indicate that aPBE0 (ML) would provide nearly exact (CCSD(T)) atomization energies if only a sufficiently predictive machine learning model of $a_{\rm opt}$ was available. 
While forces on atoms hardly change, we find that the quality of obtained electron densities is {\em not} sacrificed in favour of more accurate energetics, as it is the case for many modern DFAs in Ref.~\cite{strayingperdew}.
In particular, while the machine learning model of $a$ was regressed solely using atomization energy information of amons, the deviation of the aPBE0 based electron density from corresponding CCSD estimates also improves for the 50 QM9 molecules, down from 0.18\% for PBE0 to 0.12\% for aPBE0 for amon training set size growing from 100 to 1'169 (Fig 3B). 
The deviation decreases systematically up to a training set size of 800, suggesting that 
we have reached either the maximal extrapolative scalability of the amon training set, 
or the maximal intrinsic flexibility of the single particle picture,
or the inherent limits in the PBE approximation to the correlation potential. 
In comparison, note that conventional supervised ML models attempting to directly learn and predict electron densities struggle to reach less than 1\% deviation from reference,
even after training on up to an order of magnitude larger training set sizes.~\cite{grisafi2018transferable,fabrizio2019electron,lv2023deep}

In order to further assess the applicability of aPBE0, 
we  considered the prediction of the GW~\cite{gw} based HOMO-LUMO gaps for the first 100 small organic molecules from the QM7b\cite{CM_and_qm7,qm7b} dataset.
Typically, optimal $a$ values for organic amons increase (see Fig 1),
causing a widening of the gap, and bringing the prediction closer to the GW reference numbers.
Results in Fig 3D quantify the improvement, 
the deviation from GW is reduced from 3.52 eV for PBE0 down to 0.86 eV for aPBE0. 
aPBE0  outperforms all of the other common functionals except for M06-2X which is just 0.2 eV closer to GW on average. Note that M06-2X is a more expensive meta-GGA hybrid, but it could also be made adaptive.
Table 1 in the SI provides errors for HOMO and LUMO eigenvalues separately, showing that the improvement with aPBE0 comes exclusively from improved HOMO eigenvalue predictions.
We reiterate again, that also these improvements, just as for the electron density, 
were found by regressing on $a_{\rm opt}$ values obtained solely by optimizing the DFA atomization energy, 
i.e.~without explicitly including any  specific information about MO eigenvalues in the loss-function. 
Furthermore, this kind of transferability appears particularly robust since all the predictions presented were obtained not only for different labels (atomization energies, densities and gaps) but also for molecules drawn from   distributions (QM9 and QM7b molecules for predictions, respectively) different from those used for training (ML models were trained on $a_{\rm opt}$-values that minimize atomization energy errors for 1'169 amon based molecular fragments with no more than 5 heavy atoms). 

\section*{Revised QM9 dataset}
\begin{figure*}
    \centering
    \includegraphics[width=\linewidth]{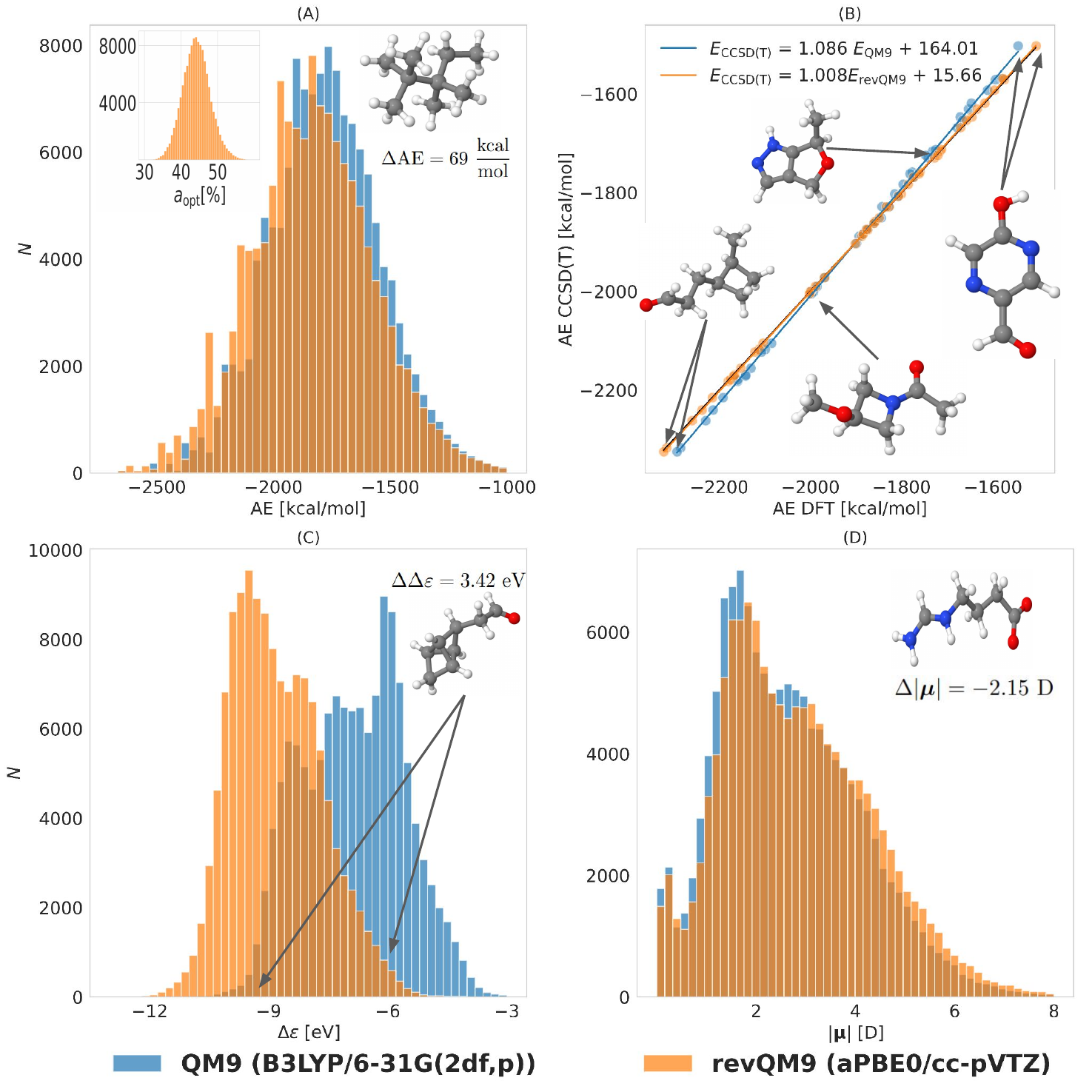}
    \caption{Changes in calculated property distributions going from QM9 (B3LYP/6-31G(2df,p)) to revQM9 (aPBE0/cc-pVTZ). 
    (A) Atomization energies bind stronger. Inset figure shows ML-predicted $a_{\mathrm{opt}}$ value distribution for all QM9 molecules, centred at 44\%, and used to generate revQM9.
    (B) Scatter plot indicates improved correlation (true diagonal shown in black) with respect to CCSD(T)/cc-pVTZ values obtained for a validation set of 50 random molecules.
    Deviation for B3LYP energies becomes larger with increasing energies much faster than for aPBE0 as apparent from the slopes.
    (C) HOMO-LUMO gaps widen. (D) Dipole moment norms increase.
    Molecules showing the largest change, along with the value (QM9-revQM9), between the two datasets are shown as insets in A, B and C.}
    \label{fig:revqm9}
\end{figure*}
The  hybrid DFA (B3LYP) based QM9 data-set~\cite{QM9} was the first to report QM properties for more than a hundred thousand systematically generated organic molecules (drawn from GDB-17~\cite{GDB17}),  
and has by now become a frequently used benchmark set for many of the popular statistical supervised learning models  which have emerged over the past decade.~\cite{huang2021abinitio} 
Already in 2017, Faber \textit{et al}.~\cite{felix_google} demonstrated that ML models can surpass hybrid DFT accuracy for all the QM properties reported in QM9, highlighting the need for more accurate higher level theory labels with chemical accuracy. 
This need was partially met by two independent efforts, both reporting expensive G4MP2 composite method based estimates of atomization energies for all of QM9.~\cite{narayanan2019qm9G4MP2,kim2019qm9G4MP2}
We have made use of the aPBE0 model described within the previous section (i.e. trained on optimal admixture ratios reproducing CCSD(T) atomization energies of 1169 organic amon fragments containing not more than 5 heavy atoms) in order to estimate $a_{\rm opt}$ for all QM9 molecules. 
Our subsequent aPBE0 calculations provide not only CCSD(T) quality estimates of energies but also properties --- at the same hybrid DFA cost as the original dataset, with stronger bonding (Fig 4A), markedly larger band-gaps (Fig 4C) and enhanced polarization (Fig 4D). 
Going beyond just energy estimates, our revised QM9 (revQM9) dataset (link provided below) includes total energies, atomization energies, all MO energies, dipole moments and density matrices.
The predicted $a_{\mathrm{opt}}$ values for the $\sim$130k molecules are shown as an inset in figure 4A and follow a nearly perfect normal distribution centered at 44\% exact exchange, with a standard deviation of only 3.7\%. 
Out of all, 8625 molecules required more than 50\% and only 6 molecules required less than 30\% exact exchange.
Computational details and timings are mentioned in the SI, we note that all the aPBE0 calculations did not consume more compute cycles than any other global hybrid DFA would have required. 
 
The distribution of aPBE0 based atomization energies is shifted on average by $\sim$16.8 kcal/mol towards stronger binding than for B3LYP. 
This large difference exceeds what one would expect from the functional alone (Fig 3B), and is likely due to a combined effect of functional and basis set: For revQM9, we have relied on the larger (triple zeta) and more balanced cc-pVTZ basis set, while the original QM9 data set was calculated with B3LYP in the 6-31G(2df,p) (double zeta) basis.
For validation purposes, we performed additional CCSD(T)/cc-pVTZ reference calculations for 50  molecules drawn at random from QM9. 
The resulting correlation between the corresponding CCSD(T) atomization energies and the B3LYP/aPBE0 values from the QM9/revQM9 datasets is also shown in figure 4B. 
Linear fits yield slopes close to one and MAEs of 15.86 and 1.98 kcal/mol, respectively.
For HOMO-LUMO gaps we found an average difference of -1.95 eV (revQM9 - QM9) between the two datasets.
Comparison with Fig 3C, and taking into account the known trend of HOMO-LUMO gaps being underestimated by KS-DFT,\cite{wires} we expect an average improvement of $\sim$ 2 eV per molecule to be likely.
We also note that our average gaps ($\sim$7 eV) and  optimal admixture values ($\sim$44 \%) in the organic molecules in revQM9 are similar to the values obtained for insulating solids reported in Ref.~\cite{yang2023range}.

For dipole moments, hybrid functionals such as B3LYP and PBE0 have been shown to be highly accurate compared to higher level wavefunction based methods\cite{dipmom_polar_benchmark,dip_mom_dft_headgord} such as CCSD. 
The size of the basis set has been shown to have a larger effect on dipole moments and polarizabilities.\cite{dipmom_polar_benchmark} 
The mean difference (revQM9 - QM9) between the dipole moment norm values is 0.18 Debye which is likely due to the better basis set employed in our calculations.
Given the encouraging results for our test set of 50 QM9 molecules in figure 3B
(the aPBE0 based $\%$ MAD to CCSD densities being 0.11 $\%$, an improvement over PBE0 (0.18) and B3LYP (0.25) (with cc-pVTZ basis set in all cases)), revQM9 also includes density matrices for all molecules which were not included in the original QM9.

\section*{Discussion and conclusions}
We have confirmed the hypothesis that system-specific optimal exact exchange admixture $a_{\rm opt}$  can be modeled by machine learning and enable a dramatic reduction in prediction errors of hybrid DFAs by reducing delocalization errors. 
Our learning curves also indicate that $a_{\rm opt}$ is sufficiently smooth and low-dimensional to enable efficient regression, and to effectively generalize in chemical compound space  with negligible computational overhead. 
Adaptive hybrid DFA (aPBE0) based calculations using `on-the-fly' machine learning based predictions of $a_{\rm opt}$  result in much improved singlet and triplet energy estimates for $\sim$ 3000 carbenes, suggesting that the spin-gap prediction problem of common DFAs could be resolved. 
aPBE0 based estimates of atomization energies, electron densities, and
HOMO-LUMO gaps of thousands of small organic molecules  also improve
significantly. 
Encouraged by the performance of aPBE0, we have revised the entire
QM9 data set, and present  improved estimates of 
energies, orbital energies, dipole moments, and electron density matrices. 

The improvement of other properties that were not part of ML training 
highlights the usefulness of adaptive parameters in Hamiltonians, 
and might even amount to a `foundation model' for quantum chemistry.
We  note that adaptive exchange and correlation do not restrict subsequent combination with other popular corrections, e.g.
enabling the simulation of non-covalent interactions, 
such as the many-body dispersion,~\cite{MBD} the vdW-DF functionals,\cite{dion2004van, klimevs2009chemical} Johnson's exchange-hole dipole moment,~\cite{xdm-chapter} or Grimme's D-series (ex. D3)~\cite{d3} method. 
Other possible improvements for specific observables are likely achievable through 
combination with $\Delta$-ML~\cite{Ramakrishnan2015, bing_DMC} and multi-level learning.~\cite{zaspel2018boosting,m3l}

While this work has focused on the HF exchange admixture ratio within the global hybrid GGA PBE0 functional due to its simplicity, clearly the same strategy could have been applied to any other hybrid functional as well.
Specifically, even lower offsets on performance curves may be achievable
by using adaptive meta-GGA/range-separated hybrids or double hybrids with adaptive correlation.
More generally speaking, we believe that there is much merit in adaptifying model Hamiltonian parameters, as also corroborated by promising results within semi-empirical quantum chemistry~\cite{Pavlo2015parameterML} or tight binding DFT~\cite{elstner2018unsupervisedTightBinding,Tkatchenko2020repulsionTightBindingDFT}.
More recent related contributions include the ML-$\omega$PBE by Lin and co-workers~\cite{ju2021stacked,ju2023accurate} 
and the ML based prediction of element-specific atom centred correction parameters by
Di Labio and co-workers~\cite{prasad2024bridging}.
Further refinements to this methodology could also be made in the spirit of Ref.~\cite{dm21}, namely by imposing additional exact constraints which can be satisfied via adaptive functional parameters that change value depending on the external potential.
To increase transferability, the functional parameters can also be optimized via variational minimization of the energy while constrained to the reference electron density.
This would ensure the simultaneous improvement of all observables.

Going beyond mere navigation of chemical space, accurate predictions of
bond dissociation profiles would enable studying the dynamics of reactive species. 
For bond dissociation, adapting the exchange admixture ratio within a global hybrid is unlikely to be sufficient due to the incorrect asymptotic behaviour.\cite{becke50years}
In particular, in the SI Fig. 5, we show dissociation energy curves for H$_2^+$, H$_2$ and N$_2$ diatomics. While trivial for H$_2^+$, due to SCF convergence problems, we did not manage to obtain optimal $a_{\rm opt}$ values for H$_2$ and N$_2$. 
Consequently, fixing the exchange term to 100\% HF, and adapting the correlation is a possible strategy, especially when going beyond the Coulson-Fischer point where strong correlation effects set in.~\cite{coulson-fischer}
Results for  smoothly varying scaling of the PBE correlation potential are encouraging 
since they recover the exact dissociation limits for all the three aforementioned cases (SI Fig 5). 
Future work will deal with all these avenues, and other possible limitations, for example due to the fixed functional form of given approximate Hamiltonians (even though adaptive force-field parameters of simple Stillinger Weber potentials indicate impressive predictive power when `learnt-on-the-fly'~\cite{lotf2004}) or due to discontinuities of adaptive parameters in certain linear combinations of configurational and compositional degrees of freedom. 

\section*{Data and code availability}
Scripts, a trained model for predicting the optimum exact exchange admixture ratio with the aPBE0 functional, generated CCSD(T) training/test data, the QMspin test set used in figure 2 and the QM7b test set used in figure 3 are available at https://github.com/dkhan42/aPBE0
\\
The revised QM9 (revQM9) dataset calculated with the aPBE0 functional and cc-pVTZ basis set is publicly available at https://doi.org/10.5281/zenodo.10689884

\bibliography{scibib}

\begin{thebibliography}{100}

\bibitem{back2024accelerated}
S.~Back, {\it et~al.\/}, {\it Digital Discovery\/} {\bf 3}, 23 (2024).

\bibitem{ceder1998predicting}
G.~Ceder, {\it Science\/} {\bf 280}, 1099 (1998).

\bibitem{franceschetti1999inverse}
A.~Franceschetti, A.~Zunger, {\it Nature\/} {\bf 402}, 60 (1999).

\bibitem{zunger2018inverse}
A.~Zunger, {\it Nature Reviews Chemistry\/} {\bf 2}, 0121 (2018).

\bibitem{smallmolinverse}
B.~Sanchez-Lengeling, A.~Aspuru-Guzik, {\it Science\/} {\bf 361}, 360 (2018).

\bibitem{carleo2017solving}
G.~Carleo, M.~Troyer, {\it Science\/} {\bf 355}, 602 (2017).

\bibitem{ferminet}
D.~Pfau, J.~S. Spencer, A.~G. Matthews, W.~M.~C. Foulkes, {\it Physical Review Research\/} {\bf 2}, 033429 (2020).

\bibitem{paulinet}
J.~Hermann, Z.~Sch{\"a}tzle, F.~No{\'e}, {\it Nature Chemistry\/} {\bf 12}, 891 (2020).

\bibitem{mattsson2002pursuit}
A.~E. Mattsson, {\it Science\/} {\bf 298}, 759 (2002).

\bibitem{lejaeghere2016reproducibility}
K.~Lejaeghere, {\it et~al.\/}, {\it Science\/} {\bf 351}, aad3000 (2016).

\bibitem{dftai}
B.~Huang, G.~F. von Rudorff, O.~A. von Lilienfeld, {\it Science\/} {\bf 381}, 170 (2023).

\bibitem{strayingperdew}
M.~G. Medvedev, I.~S. Bushmarinov, J.~Sun, J.~P. Perdew, K.~A. Lyssenko, {\it Science\/} {\bf 355}, 49 (2017).

\bibitem{burke2012JCP}
K.~Burke, {\it J. Chem. Phys.\/} {\bf 136}, 150901 (2012).

\bibitem{becke50years}
A.~D. Becke, {\it The Journal of chemical physics\/} {\bf 140} (2014).

\bibitem{cohen2008insights}
A.~J. Cohen, P.~Mori-S{\'a}nchez, W.~Yang, {\it Science\/} {\bf 321}, 792 (2008).

\bibitem{wires}
K.~R. Bryenton, A.~A. Adeleke, S.~G. Dale, E.~R. Johnson, {\it Wiley Interdiscip. Rev. Comput. Mol. Sci.\/} p. e1631 (2022).

\bibitem{sham1985density}
L.~Sham, M.~Schl{\"u}ter, {\it Physical Review B\/} {\bf 32}, 3883 (1985).

\bibitem{perdew1985density}
J.~P. Perdew, {\it International Journal of Quantum Chemistry\/} {\bf 28}, 497 (1985).

\bibitem{cohen2008fractional}
A.~J. Cohen, P.~Mori-S{\'a}nchez, W.~Yang, {\it Physical Review B\/} {\bf 77}, 115123 (2008).

\bibitem{hostavs2023important}
J.~Hosta{\v{s}}, {\it et~al.\/}, {\it The Journal of Chemical Physics\/} {\bf 159} (2023).

\bibitem{burke1997adiabatic}
K.~Burke, M.~Ernzerhof, J.~P. Perdew, {\it Chemical Physics Letters\/} {\bf 265}, 115 (1997).

\bibitem{b3}
A.~D. Becke, {\it J. Chem. Phys.\/} {\bf 98}, 5648 (1993).

\bibitem{pbe0}
C.~Adamo, V.~Barone, {\it J. Chem. Phys.\/} {\bf 110}, 6158 (1999).

\bibitem{PBE01}
M.~Ernzerhof, G.~E. Scuseria, {\it J. Comp. Phys.\/} {\bf 110}, 5029 (1999).

\bibitem{brorsen2017accuracy}
K.~R. Brorsen, Y.~Yang, M.~V. Pak, S.~Hammes-Schiffer, {\it The journal of physical chemistry letters\/} {\bf 8}, 2076 (2017).

\bibitem{langreth1975exchange}
D.~C. Langreth, J.~P. Perdew, {\it Solid State Communications\/} {\bf 17}, 1425 (1975).

\bibitem{poison}
T.~Gould, S.~G. Dale, {\it Physical Chemistry Chemical Physics\/} {\bf 24}, 6398 (2022).

\bibitem{price2023xdm}
A.~J. Price, A.~Otero-de-la Roza, E.~R. Johnson, {\it Chemical Science\/} {\bf 14}, 1252 (2023).

\bibitem{price2023accurate}
A.~J. Price, R.~A. Mayo, A.~Otero-de-la Roza, E.~R. Johnson, {\it CrystEngComm\/} {\bf 25}, 953 (2023).

\bibitem{cocrystal_erin2018}
L.~M. LeBlanc, {\it et~al.\/}, {\it Angewandte Chemie International Edition\/} {\bf 57}, 14906 (2018).

\bibitem{ruiz1996charge}
E.~Ruiz, D.~R. Salahub, A.~Vela, {\it The Journal of Physical Chemistry\/} {\bf 100}, 12265 (1996).

\bibitem{sini2011evaluating}
G.~Sini, J.~S. Sears, J.-L. Brédas, {\it Journal of Chemical Theory and Computation\/} {\bf 7}, 602 (2011). PMID: 26596294.

\bibitem{steinmann2012interaction}
S.~N. Steinmann, C.~Piemontesi, A.~Delachat, C.~Corminboeuf, {\it Journal of Chemical Theory and Computation\/} {\bf 8}, 1629 (2012).

\bibitem{cai2002failure}
Z.-L. Cai, K.~Sendt, J.~R. Reimers, {\it The Journal of chemical physics\/} {\bf 117}, 5543 (2002).

\bibitem{tozer2003relationship}
D.~J. Tozer, {\it The Journal of chemical physics\/} {\bf 119}, 12697 (2003).

\bibitem{dreuw2003long}
A.~Dreuw, J.~L. Weisman, M.~Head-Gordon, {\it The Journal of chemical physics\/} {\bf 119}, 2943 (2003).

\bibitem{otero2014halogen}
A.~Otero-De-La-Roza, E.~R. Johnson, G.~A. DiLabio, {\it Journal of chemical theory and computation\/} {\bf 10}, 5436 (2014).

\bibitem{lynch2001well}
B.~J. Lynch, D.~G. Truhlar, {\it The Journal of Physical Chemistry A\/} {\bf 105}, 2936 (2001).

\bibitem{janesko2008hartree}
B.~G. Janesko, G.~E. Scuseria, {\it The Journal of chemical physics\/} {\bf 128} (2008).

\bibitem{nandy2020large}
A.~Nandy, {\it et~al.\/}, {\it Physical Chemistry Chemical Physics\/} {\bf 22}, 19326 (2020).

\bibitem{zhao2019stable}
Q.~Zhao, H.~J. Kulik, {\it The journal of physical chemistry letters\/} {\bf 10}, 5090 (2019).

\bibitem{santra2021types}
G.~Santra, J.~M. Martin, {\it Journal of chemical theory and computation\/} {\bf 17}, 1368 (2021).

\bibitem{gmtkn55}
L.~Goerigk, {\it et~al.\/}, {\it Phys. Chem. Chem. Phys.\/} {\bf 19}, 32184 (2017).

\bibitem{yang2023range}
J.~Yang, S.~Falletta, A.~Pasquarello, {\it npj Computational Materials\/} {\bf 9}, 108 (2023).

\bibitem{friede2023optimally}
M.~Friede, S.~Ehlert, S.~Grimme, J.-M. Mewes, {\it Journal of Chemical Theory and Computation\/} {\bf 19}, 8097 (2023).

\bibitem{CM_and_qm7}
M.~Rupp, A.~Tkatchenko, K.-R. M\"uller, O.~A. von Lilienfeld, {\it Phys. Rev. Lett.\/} {\bf 108}, 058301 (2012).

\bibitem{huang2021abinitio}
B.~Huang, O.~A. von Lilienfeld {\bf 121}, 10001 (2021).

\bibitem{mbdf}
D.~Khan, S.~Heinen, O.~A. von Lilienfeld, {\it The Journal of Chemical Physics\/} {\bf 159}, 034106 (2023).

\bibitem{Brockherde2017}
F.~Brockherde, {\it et~al.\/}, {\it Nature Communications\/} {\bf 8} (2017).

\bibitem{dick2020machine}
S.~Dick, M.~Fernandez-Serra, {\it Nature communications\/} {\bf 11}, 3509 (2020).

\bibitem{nagai2020completing}
R.~Nagai, R.~Akashi, O.~Sugino, {\it npj Computational Materials\/} {\bf 6}, 43 (2020).

\bibitem{dm21}
J.~Kirkpatrick, {\it et~al.\/}, {\it Science\/} {\bf 374}, 1385 (2021).

\bibitem{Margraf2021}
J.~T. Margraf, K.~Reuter, {\it Nature Communications\/} {\bf 12} (2021).

\bibitem{welborn2018transferability}
M.~Welborn, L.~Cheng, T.~F. Miller~III, {\it Journal of chemical theory and computation\/} {\bf 14}, 4772 (2018).

\bibitem{karandashev2022orbital}
K.~Karandashev, O.~A. von Lilienfeld, {\it The Journal of Chemical Physics\/} {\bf 156} (2022).

\bibitem{Ramakrishnan2015}
R.~Ramakrishnan, P.~O. Dral, M.~Rupp, O.~A. von Lilienfeld, {\it Journal of Chemical Theory and Computation\/} {\bf 11}, 2087–2096 (2015).

\bibitem{Bogojeski2020}
M.~Bogojeski, L.~Vogt-Maranto, M.~E. Tuckerman, K.-R. M\"{u}ller, K.~Burke, {\it Nature Communications\/} {\bf 11} (2020).

\bibitem{pribram2015dft}
A.~Pribram-Jones, D.~A. Gross, K.~Burke, {\it Annual review of physical chemistry\/} {\bf 66}, 283 (2015).

\bibitem{pbe}
J.~P. Perdew, K.~Burke, M.~Ernzerhof, {\it Phys. Rev. Lett.\/} {\bf 77}, 3865 (1996).

\bibitem{QM9}
R.~Ramakrishnan, P.~O. Dral, M.~Rupp, O.~A. Von~Lilienfeld, {\it Scientific data\/} {\bf 1}, 1 (2014).

\bibitem{GDB17}
L.~Ruddigkeit, R.~van Deursen, L.~Blum, J.-L. Reymond {\bf 52}, 2684 (2012).

\bibitem{qmspin}
M.~Schwilk, D.~N. Tahchieva, O.~A. von Lilienfeld, {\it arXiv preprint arXiv:2004.10600\/}  (2020).

\bibitem{vapnik1994learningcurves}
C.~Cortes, L.~D. Jackel, S.~A. Solla, V.~Vapnik, J.~S. Denker, {\it Advances in Neural Information Processing Systems\/} (1993), pp. 327--334.

\bibitem{amons}
B.~Huang, O.~A. von Lilienfeld, {\it Nature chemistry\/} {\bf 12}, 945 (2020).

\bibitem{bing_DMC}
B.~Huang, O.~A. von Lilienfeld, J.~T. Krogel, A.~Benali, {\it Journal of Chemical Theory and Computation\/} {\bf 19}, 1711 (2023).

\bibitem{gw}
L.~Hedin, {\it Physical Review\/} {\bf 139}, A796 (1965).

\bibitem{qm7b}
G.~Montavon, {\it et~al.\/}, {\it New Journal of Physics\/} {\bf 15}, 095003 (2013).

\bibitem{grisafi2018transferable}
A.~Grisafi, {\it et~al.\/}, {\it ACS central science\/} {\bf 5}, 57 (2018).

\bibitem{fabrizio2019electron}
A.~Fabrizio, A.~Grisafi, B.~Meyer, M.~Ceriotti, C.~Corminboeuf, {\it Chemical science\/} {\bf 10}, 9424 (2019).

\bibitem{lv2023deep}
T.~Lv, {\it et~al.\/}, {\it Physical Review B\/} {\bf 108}, 235159 (2023).

\bibitem{felix_google}
F.~A. Faber, {\it et~al.\/}, {\it Journal of Chemical Theory and Computation\/} {\bf 13}, 5255 (2017). PMID: 28926232.

\bibitem{narayanan2019qm9G4MP2}
B.~Narayanan, P.~C. Redfern, R.~S. Assary, L.~A. Curtiss, {\it Chemical science\/} {\bf 10}, 7449 (2019).

\bibitem{kim2019qm9G4MP2}
H.~Kim, J.~Y. Park, S.~Choi, {\it Scientific data\/} {\bf 6}, 109 (2019).

\bibitem{dipmom_polar_benchmark}
A.~L. Hickey, C.~N. Rowley, {\it The Journal of Physical Chemistry A\/} {\bf 118}, 3678 (2014).

\bibitem{dip_mom_dft_headgord}
D.~Hait, M.~Head-Gordon, {\it Journal of chemical theory and computation\/} {\bf 14}, 1969 (2018).

\bibitem{MBD}
A.~Tkatchenko, {R. A. DiStasio Jr.}, R.~Car, M.~Scheffler, {\it Phys. Rev. Lett.\/} {\bf 108}, 236402 (2012).

\bibitem{dion2004van}
M.~Dion, H.~Rydberg, E.~Schr{\"o}der, D.~C. Langreth, B.~I. Lundqvist, {\it Physical review letters\/} {\bf 92}, 246401 (2004).

\bibitem{klimevs2009chemical}
J.~Klime{\v{s}}, D.~R. Bowler, A.~Michaelides, {\it Journal of Physics: Condensed Matter\/} {\bf 22}, 022201 (2009).

\bibitem{xdm-chapter}
E.~R. Johnson, {\it Non-covalent Interactions in Quantum Chemistry and Physics\/}, A.~{Otero-de-la-Roza}, G.~A. DiLabio, eds. (Elsevier, 2017), chap.~5, pp. 169--194.

\bibitem{d3}
S.~Grimme, J.~Antony, S.~Ehrlich, H.~Krieg, {\it J. Chem. Phys.\/} {\bf 132}, 154104 (2010).

\bibitem{zaspel2018boosting}
P.~Zaspel, B.~Huang, H.~Harbrecht, O.~A. von Lilienfeld, {\it Journal of chemical theory and computation\/} {\bf 15}, 1546 (2018).

\bibitem{m3l}
S.~Heinen, {\it et~al.\/}, {\it arXiv preprint arXiv:2308.11196\/}  (2023).

\bibitem{Pavlo2015parameterML}
P.~O. Dral, O.~A. von Lilienfeld, W.~Thiel, {\it Journal of Chemical Theory and Computation\/} {\bf 11}, 2120 (2015). PMID: 26146493.

\bibitem{elstner2018unsupervisedTightBinding}
J.~J. Kranz, M.~Kubillus, R.~Ramakrishnan, O.~A. von Lilienfeld, M.~Elstner, {\it Journal of chemical theory and computation\/} {\bf 14}, 2341 (2018).

\bibitem{Tkatchenko2020repulsionTightBindingDFT}
M.~St\"ohr, L.~Medrano~Sandonas, A.~Tkatchenko, {\it The Journal of Physical Chemistry Letters\/} {\bf 11}, 6835 (2020).

\bibitem{ju2021stacked}
C.-W. Ju, E.~J. French, N.~Geva, A.~W. Kohn, Z.~Lin, {\it The Journal of Physical Chemistry Letters\/} {\bf 12}, 9516 (2021).

\bibitem{ju2023accurate}
C.-W. Ju, {\it et~al.\/}, {\it The Journal of Physical Chemistry A\/}  (2023).

\bibitem{prasad2024bridging}
V.~K. Prasad, A.~Otero-de-la Roza, G.~diLabio, {\it Machine Learning: Science and Technology\/} {\bf 5}, 015035 (2024).

\bibitem{coulson-fischer}
C.~Coulson, I.~Fischer, {\it The London, Edinburgh, and Dublin Philosophical Magazine and Journal of Science\/} {\bf 40}, 386 (1949).

\bibitem{lotf2004}
G.~Cs{\'a}nyi, T.~Albaret, M.~C. Payne, A.~D. Vita, {\it Phys. Rev. Lett.\/} {\bf 93}, 175503 (2004).

\bibitem{perdew1996rationale}
J.~P. Perdew, M.~Ernzerhof, K.~Burke, {\it The Journal of chemical physics\/} {\bf 105}, 9982 (1996).

\bibitem{vcivzek1966correlation}
J.~{\v{C}}{\'\i}{\v{z}}ek, {\it The Journal of Chemical Physics\/} {\bf 45}, 4256 (1966).

\bibitem{vcivzek1971correlation}
J.~{\v{C}}{\'\i}{\v{z}}ek, J.~Paldus, {\it International Journal of Quantum Chemistry\/} {\bf 5}, 359 (1971).

\bibitem{raghavachari1989fifth}
K.~Raghavachari, G.~W. Trucks, J.~A. Pople, M.~Head-Gordon, {\it Chemical Physics Letters\/} {\bf 157}, 479 (1989).

\bibitem{scuseria1990comparison}
G.~E. Scuseria, T.~J. Lee, {\it The Journal of chemical physics\/} {\bf 93}, 5851 (1990).

\bibitem{werner1988efficient}
H.-J. Werner, P.~J. Knowles, {\it The Journal of chemical physics\/} {\bf 89}, 5803 (1988).

\bibitem{KNOWLES1988514}
P.~J. Knowles, H.-J. Werner, {\it Chemical Physics Letters\/} {\bf 145}, 514 (1988).

\bibitem{mrci-f12}
T.~Shiozaki, G.~Knizia, H.-J. Werner, {\it The Journal of Chemical Physics\/} {\bf 134}, 034113 (2011).

\bibitem{mrci-f12_2}
T.~Shiozaki, H.-J. Werner, {\it The Journal of Chemical Physics\/} {\bf 134}, 184104 (2011).

\bibitem{mrci-f12_theory}
T.~Shiozaki, H.-J. Werner, {\it Molecular Physics\/} {\bf 111}, 607 (2013).

\bibitem{ccpvtz}
R.~A. Kendall, T.~H. Dunning~Jr, R.~J. Harrison, {\it The Journal of chemical physics\/} {\bf 96}, 6796 (1992).

\bibitem{sun2020recent}
Q.~Sun, {\it et~al.\/}, {\it The Journal of chemical physics\/} {\bf 153} (2020).

\bibitem{sun2018pyscf}
Q.~Sun, {\it et~al.\/}, {\it Wiley Interdisciplinary Reviews: Computational Molecular Science\/} {\bf 8}, e1340 (2018).

\bibitem{sun2015libcint}
Q.~Sun, {\it Journal of computational chemistry\/} {\bf 36}, 1664 (2015).

\bibitem{lyp}
C.~Lee, W.~Yang, R.~G. Parr, {\it Phys. Rev. B\/} {\bf 37}, 785 (1988).

\bibitem{def2tzvp}
F.~Weigend, R.~Ahlrichs, {\it Physical Chemistry Chemical Physics\/} {\bf 7}, 3297 (2005).

\bibitem{chen2023physics}
K.~Chen, C.~Kunkel, B.~Cheng, K.~Reuter, J.~T. Margraf, {\it Chemical Science\/} {\bf 14}, 4913 (2023).

\bibitem{cignoni2023electronic}
E.~Cignoni, {\it et~al.\/}, Electronic excited states from physically-constrained machine learning (2023).

\bibitem{vapnik1999nature}
V.~Vapnik, {\it The nature of statistical learning theory\/} (Springer science \& business media, 1999).

\bibitem{rasmussen2006gaussian}
C.~E. Rasmussen, {\it et~al.\/}, {\it Gaussian processes for machine learning\/}, vol.~1 (Springer).

\bibitem{GPR_deringer}
V.~L. Deringer, {\it et~al.\/}, {\it Chemical Reviews\/} {\bf 121}, 10073 (2021). PMID: 34398616.

\bibitem{siwoo}
S.~Lee, S.~Heinen, D.~Khan, O.~A. von Lilienfeld, {\it arXiv preprint arXiv:2308.00389\/}  (2023).

\bibitem{fchl19}
A.~S. Christensen, L.~A. Bratholm, F.~A. Faber, O.~Anatole~von Lilienfeld, {\it The Journal of Chemical Physics\/} {\bf 152}, 044107 (2020).

\bibitem{excited_states_ceriotti}
E.~Cignoni, {\it et~al.\/}, {\it arXiv preprint arXiv:2311.00844\/}  (2023).

\bibitem{behler_HDNN}
J.~Behler, M.~Parrinello, {\it Phys. Rev. Lett.\/} {\bf 98}, 146401 (2007).

\bibitem{bob}
K.~Hansen, {\it et~al.\/}, {\it The journal of physical chemistry letters\/} {\bf 6}, 2326 (2015).

\bibitem{braida2011quantum}
B.~Bra{\"\i}da, J.~Toulouse, M.~Caffarel, C.~Umrigar, {\it The Journal of chemical physics\/} {\bf 134} (2011).

\bibitem{morse_N2_alastair}
B.~Braïda, J.~Toulouse, M.~Caffarel, C.~J. Umrigar, {\it The Journal of Chemical Physics\/} {\bf 134}, 084108 (2011).

\end{thebibliography}

\bibliographystyle{Science}

\section*{Acknowledgements}
We acknowledge discussions with B. Huang, E. Johnson, and assistance by O. Trottier. 
We acknowledge the support of the Natural Sciences and Engineering Research Council of Canada (NSERC), [funding reference number RGPIN-2023-04853]. Cette recherche a été financée par le Conseil de recherches en sciences naturelles et en génie du Canada (CRSNG), [numéro de référence RGPIN-2023-04853].
This research was undertaken thanks in part to funding provided to the University of Toronto's Acceleration Consortium from the Canada First Research Excellence Fund,
grant number: CFREF-2022-00042.
O.A.v.L. has received support as the Ed Clark Chair of Advanced Materials and as a Canada CIFAR AI Chair.
O.A.v.L. has received funding from the European Research Council (ERC) under the European Union’s Horizon 2020 research and innovation programme (grant agreement No. 772834).

\textbf{Author contributions} : 
D.K., A.J.A.P., and O.A.v.L. conceived the project, analyzed the results and wrote the manuscript.
D.K. designed and implemented the methods, and performed calculations. 
D.K. and M.L.A. prepared figures.

\section*{Supplementary Material} 

\begin{figure*}
    \centering
     \includegraphics[width=\linewidth]{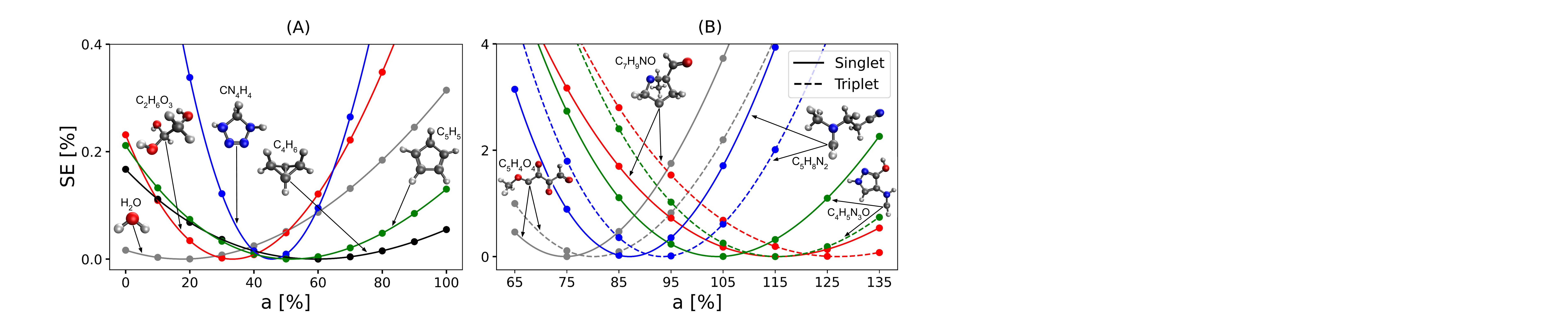}
    
    \caption{Squared atomization energy errors of PBE0 calculations with variable HF exchange percentage for (A) for small organic molecular fragments (amons\cite{amons}) reported in Ref.~\cite{bing_DMC} with upto 5 heavy atoms compared to CCSD(T) and (B) carbenes from the QMSpin\cite{qmspin} dataset compared to MRCISD+Q.
    A grid search was conducted and the optimal exact exchange ratio was found via a quartic polynomial fit. 
    The procedure was repeated and labels were generated for all molecules used during training of the machine learning models.}
    \label{fig:error_vs_X}
\end{figure*}
\subsection{Adaptive hybrids}
Hybrid exchange-correlation DFAs mix a portion of HF exchange energy ($E_{\mathrm{X}}^\mathrm{HF}$)
\\
\\
\\
\\
\begin{equation}
   \mathrm{E}_\mathrm{X}^\mathrm{HF} = -\frac{1}{2}\sum_{\sigma}\sum_{ij}\int \int \varphi_{i \sigma}^*(\mathbf{r})\varphi_{j \sigma}^*(\mathbf{r'})\frac{1}{|\mathbf{r} - \mathbf{r'}|}\varphi_{i \sigma}(\mathbf{r'})\varphi_{j \sigma}(\mathbf{r})d\mathbf{r}d\mathbf{r'},
   \label{HF_x}
\end{equation}
where $\varphi_{i \sigma}$ are spin orbitals; 
with the DFA exchange ($E_\mathrm{X}^\mathrm{DFA}$) and correlation ($E_\mathrm{C}^\mathrm{DFA}$) energy\cite{b3,perdew1996rationale} 
\begin{equation}
    E_\mathrm{XC}^\mathrm{hybrid} = a E_\mathrm{X}^\mathrm{HF} + (1-a)E_\mathrm{X}^\mathrm{DFA}+ E_\mathrm{C}^\mathrm{DFA}
    \label{hybrid}
\end{equation}
with the admixture ratio of exchange $a$ being close to 25\% for popular hybrid functionals such as PBE0\cite{pbe0,pbe} and B3LYP\cite{b3}.
In our work we assume the optimal admixture ratio to adapt, i.e. to be system specific in the sense that it becomes a function of the external potential and the electronic state, which can be machine learnt and predicted on the fly for new, out-of-sample systems. 
We dub this the adaptive PBE0 functional  (aPBE0) for varying the balance of HF and PBE exchange,
\begin{equation}
    E_\mathrm{xc}^\mathrm{aPBE0} = a(\{ Z_{I},\mathbf{R}_{I}\}_{I},{S})E_\mathrm{X}^\mathrm{HF} + (1-a(\{ Z_{I},\mathbf{R}_{I}\}_{I},{S}))E_\mathrm{X}^\mathrm{PBE} + b E_\mathrm{C}^\mathrm{PBE}
    \label{aPBE0}
\end{equation}
where $\{ Z_{I},\mathbf{R}_{I}\}_{I}$ denotes the set of nuclear charges and coordinates respectively, ${S}$ denotes the electronic spin state of the system, and the DFT exchange-correlation energies come from the PBE\cite{pbe} functional.

Conversely, when varying correlation, we set $a$ to 100\% and predict the amount of PBE correlation that is blended in. More specifically, 
\begin{equation}
    E_\mathrm{xc}^\mathrm{aPBE0c} = E_\mathrm{X}^\mathrm{HF} + b(\{ Z_{I},\mathbf{R}_{I}\}_{I},S) E_\mathrm{C}^\mathrm{PBE}
    \label{aPBE0c}
\end{equation}
The correlation term in this functional now corrects for the errors in the kinetic energy, Coulombic repulsion and HF exchange in the Kohn-Sham system.

\subsection{Optimization}
The optimal admixture ratios $a_{\mathrm{opt}}$ or $b_{\rm opt}$ were always found by minimizing the squared error of the atomization energy $E_{\mathrm{atm.}}$ with respect to a high level reference
\begin{equation}
    a_{\mathrm{opt}} = \arg \min_{a} ~(E^{\mathrm{aPBE0}}_{\mathrm{atm.}}(a) - E^{\mathrm{Ref.}}_{\mathrm{atm.}})^{2}
    \label{opt_atm}
\end{equation}
where the reference values $E^{\mathrm{Ref.}}_{\mathrm{atm.}}$ correspond to either CCSD(T)/QMC \cite{vcivzek1966correlation, vcivzek1971correlation, raghavachari1989fifth, scuseria1990comparison,bing_DMC} or Multi-Reference Configuration Interaction Singles Doubles with Davidson correction (MRCISD+Q)\cite{werner1988efficient, KNOWLES1988514, mrci-f12, mrci-f12_2, mrci-f12_theory} values in our work.
The aPBE0 atomization energies are defined as
\begin{equation}
    E^{\mathrm{aPBE0}}_{\mathrm{atm.}}(a) = E^{\mathrm{aPBE0}}(a) - \sum_{i} \varepsilon_{i}^{\mathrm{aPBE0}}(a)
    \label{atm_definition}
\end{equation}
where $E^{\mathrm{aPBE0}}(a)$ denotes the total energy and $\varepsilon_{i}^{\mathrm{aPBE0}}(a)$ denotes the aPBE0 free atom energy of atom $i$ using admixture ratio $a$.
Unless stated otherwise, the cc-pVTZ\cite{ccpvtz} basis set was employed and the PySCF 2.4.0\cite{sun2020recent,sun2018pyscf,sun2015libcint} package was used for all DFT and (all electron) CCSD(T) calculations.
For optimization we performed a simple grid scan by obtaining $E^{\mathrm{aPBE0}}_{\mathrm{atm.}}(a)$ at evenly spaced values of $a$ between 0 and 100\%.
The squared error to the reference (eq. 5) was then interpolated using a quartic polynomial fit to obtain the minimum which we found to be nearly exact in all cases.

Following this procedure we performed CCSD(T) and PBE0 calculations for the 1169 small organic molecular fragments (amons\cite{amons}) reported in Ref.~\cite{bing_DMC} to find the optimal admixture ratios for all molecules.
A few examples of these are shown in figure 1A with appreciably different $a_{\mathrm{opt}}$ values.
The distribution plot of these 1169 $a_{\mathrm{opt}}$ values is shown in figure 2A (and figure 1 in the main text) indicating a normal distribution centered at 42 $\%$ HF exchange
Throughout our work we found an $a_{\mathrm{opt}}$ value to exist which reduces the atomization energy error to nearly 0 for all cases.
The mean absolute error for these 1169 atomization energies is reduced from 3.46 to 0.02 kcal/mol by using the optimal HF exchange percentage with the PBE0 functional for all molecules.
A similar procedure was followed to obtain $a_{\mathrm{opt}}$ for $\sim$3000 carbenes in singlet and triplet state from the QMSpin dataset\cite{qmspin}.
All calculations were performed using the Restricted Open shell Kohn-Sham (ROKS) method
and all geometries used correspond to the ROKS B3LYP/def2-TZVP\cite{b3,lyp,def2tzvp} optimized triplet state structures reported in QMspin.
Separate $a_{\mathrm{opt}}$ were obtained for singlet and triplet state structures via optimization to MRCISD+Q atomization energies.
Figure 1B shows the optimization fits for a few carbenes in the singlet and triplet states while
2B shows the distribution of the obtained $a_{\mathrm{opt}}^{s}$, $a_{\mathrm{opt}}^{t}$ values for the $\sim$3000 carbenes.
Compared to their closed shell counterparts from QM9, the $a_{\mathrm{opt}}$ values for these open shell systems show a significant shift towards larger values of HF exchange with mean values being 85\% and 77\% for the singlet and triplet structures respectively.
A strong correlation can be observed between the two values with the following linear fit providing a $R^{2}$ value of $\sim$96 $\%$
 \begin{equation}
     a_{\mathrm{opt}}^{t} = 1.102a_{\mathrm{opt}}^{s} - 0.118
 \end{equation}
An important point we note here is that these $a_{\mathrm{opt}}$ values are highly sensitive to the reference atomization energies which were not reported in the QMspin dataset.
Optimization for total energies is infeasible due to the difference in reference energies between the different level of theories.
For consistency we employed the same CCSD(T) free-atom reference energies for the carbenes as well but we found a significant shift towards much smaller $a_{\mathrm{opt}}$ values ($\sim$20$\%$) when employing CISD atomic energies in this case. 
This is likely one of the reasons behind the shift to much larger values compared to Fig 2A and for values larger than 100$\%$.
However, we note that this only causes a shift in the $a_{\mathrm{opt}}$ values which were always found to exist and hence the results reported in the main text should remain valid for any reasonably accurate atomic energy reference.
Ideally one would always use the same atomic energies as the level of theory used to obtain the reference total energies, or simply rely upon experimental atomization energies.
\begin{figure*}
    \centering
    \includegraphics[width=\linewidth]{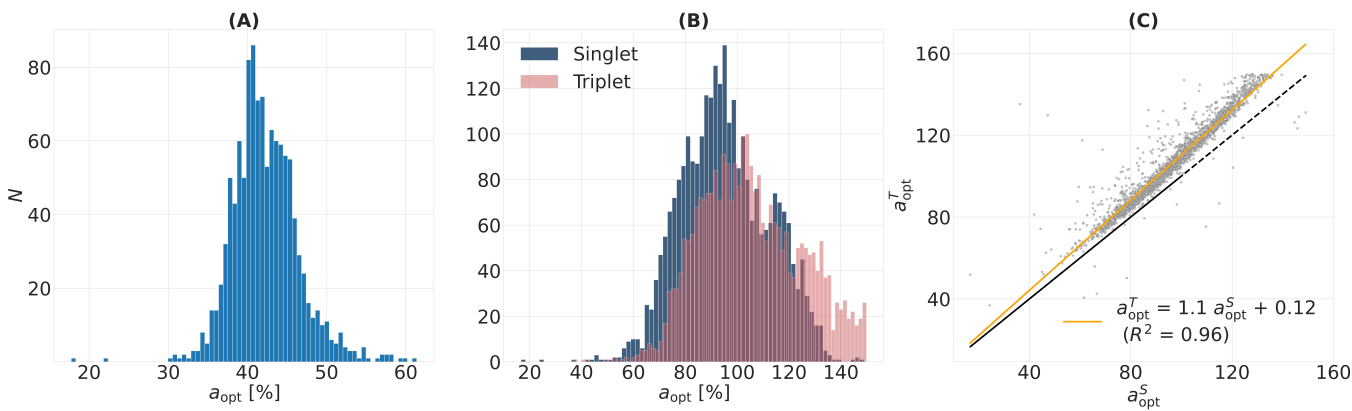}
    \caption{Distribution of optimal HF exchange $\%$, $a_{\rm opt}$, values for the generated training sets. (A) 1169 small organic molecular fragments (amons\cite{amons}) reported in Ref.~\cite{bing_DMC} obtained by optimizing to CCSD(T) atomization energies. (B) 2802 singlet ($a_{\mathrm{opt}}^{s}$) and triplet ($a_{\mathrm{opt}}^{t}$) spin states of carbenes (in their triplet geometry) obtained by optimizing to atomization energies calculated using MRCISD+Q total energies from the QMspin dataset~\cite{qmspin}. 
    CCSD(T) reference atomic energies were used to calculate atomization energies.
    (C) Corresponding correlation plot between the $a_{\mathrm{opt}}^{s}$ and $a_{\mathrm{opt}}^{t}$ values.
    }
    \label{fig:carbs_aopt_dist}
\end{figure*}
\subsection{Machine Learning \& revQM9}
This can be done efficiently by using high-level quantum reference training data for atoms-in-molecules based fragments (amons)\cite{amons} due to their small size.
Subsequent training and predictions are done on the product of $a_\mathrm{opt}$ and the number of electrons in the system which makes it an extensive property and the QML model highly transferable to larger systems thanks to local partitioning schemes\cite{bing_DMC,amons,chen2023physics,cignoni2023electronic}.

With the generated $a_{\mathrm{opt}}$ values as training data, ML models were developed to predict the optimal HF exchange values to be used alongside the PBE functional for systems of interest.
The ML model used throughout this work is Kernel Ridge Regression\cite{vapnik1999nature,rasmussen2006gaussian, GPR_deringer} (KRR) due to its high robustness and simplicity, which has been used extensively with QML models\cite{bing_DMC,felix_google,m3l,siwoo}.
The optimal HF exchange values for the query molecule, $a_{\mathrm{opt}}^{\mathrm{q}}$, can be obtained as weighted sums of similarity measures to all molecules in the training set
\begin{equation}
    a_{\mathrm{opt}}^{\mathrm{q}} = \sum_{J}^{N_{\mathrm{train}}} \alpha_{J}k(\mathbf{M}^{q}, \mathbf{M}_{J})
\end{equation}
where $\alpha_{j}$ are the regression weights, $\mathbf{M}$ are molecular representation feature vectors which depend only on the set of nuclear charges and coordinates, and $k(.,.)$ denotes a kernel function acting as a similarity measure.
The kernel function primarily used in our work is the screened local Gaussian kernel
\begin{equation}
    k(\mathbf{M}_{I}, \mathbf{M}_{J}) = \sum_{\mu\epsilon I}\sum_{\nu\epsilon J}\mathbf{\delta}_{Z_{\mu},Z_{\nu}}  \exp\left( -\frac{\vert \vert \mathbf{x}_{I\mu} - \mathbf{x}_{J\nu} \vert \vert^{2}_{2}}{2 \sigma^2} \right)~
    \label{eq:local_gaussian}
\end{equation}
where $\mathbf{x}_{I\mu}$ denotes the representation vector of atom $a$ within molecule $I$ and $\mathbf{\delta}_{Z_{\mu},Z_{\nu}} $ denotes a Kronecker Delta over the nuclear charges $Z_{\mu},Z_{\nu}$ which restricts the similarity measurement between atoms of the same chemical element\cite{fchl19}.
The regression weights $\mathbf{\alpha}$ are obtained from the set of training labels $\mathbf{y}^{\mathrm{train}}$ via the following equation

\begin{equation}
    \mathbf{\alpha} = (\mathbf{K} + \lambda \cdot \mathbf{I})^{-1} \mathbf{y}^{\mathrm{train}}
    \label{eq:solution_alpha}
\end{equation}

where $\mathbf{K}$ is the kernel matrix of the training set and $\lambda$ is a regularization parameter.
The form of the kernel function in eq. 8 partitions the system into atomic contributions.
This local partitioning is vital to the transferability of ML models to larger systems and has been successfully applied to the learning of size extensive properties\cite{amons,excited_states_ceriotti,behler_HDNN}.
However, such partitioning is not suitable to the learning of intensive properties such as the HF exchange admixture ratio, $a_{\mathrm{opt}}$, in our case.\cite{felix_google}
To circumvent this issue we follow the recipe of Chen et al~\cite{chen2023physics}
to instead learn the product of the intensive property with the extensive
number of particles. More specifically, we predict the product of $a_{\mathrm{opt}}$ with the total number of electrons ($N_{e}$) with an atomic ML model. 
The label $y_{\mathrm{opt}}$ to be learned and predicted then becomes
\begin{equation}
    y_{\mathrm{opt}} = N_{e}a_{\mathrm{opt}}
    \label{eq:intensive_extensive}
\end{equation}
\\
The representation used for generating atomic feature vectors for the ML model in our work is an extended version of the Many-Body Distribution Functionals (MBDF) representation\cite{mbdf}.
The reason for this choice is its compact size, leading to fast predictions, but nevertheless high predictive power.
Briefly, MBDF uses functionals of two- and three-body distribution functionals as atomic feature vector components:
\begin{equation}
    F_{2}^{nm}[i]=\int_{0}^{\infty}dr~g_{n}(r) ~\partial_{r}^{m}\rho_{i}(r),~
    \rho_{i}(r) = \sum_{j}^{M} ~Z_{j}\mathcal{N}{\left(R_{ij},\sigma_{r}\right)}
    \nonumber
\end{equation}
\begin{equation}
    F_{3}^{nm}[i] =\int_{0}^{\pi}d\theta~g_{n}(\theta) ~\partial_{\theta}^{m}\rho_{i}(\theta),~
    \rho_{i}(\theta) = \sum_{jk}^{M} ~\frac{(Z_{j}Z_{k})^\frac{1}{2}\mathcal{N}\left(\theta_{ijk},\sigma_\theta\right)}{(R_{ij}R_{jk}R_{ik})^2}
    \nonumber
\end{equation}
where $\mathcal{N}$ denotes a normalized Gaussian, $Z_{j}$ denotes atomic number of atom $j$, $R_{ij}$, $\theta_{ijk}$ denote interatomic distances and angles respectively and $g_{n}$ are suitable weighting functions similar to 2, 3-body interaction potentials.
\\
DFT calculations for generating the revQM9 dataset were performed with the aPBE0 functional and the cc-pVTZ basis set using the PySCF 2.4.0 package~\cite{sun2018pyscf} for $\sim$130k QM9 molecules.
The $a_{\rm opt}$ values for all molecules were predicted by the ML model as described in the main text.
The calculations required $\sim$5 days parallelized over 3 in-house nodes (36 core 4.8GHz Intel Xeon W9-3475X/1 TB DDR5 ECC RAM).

\begin{figure}
    \centering 
    \includegraphics[width=\columnwidth]
    {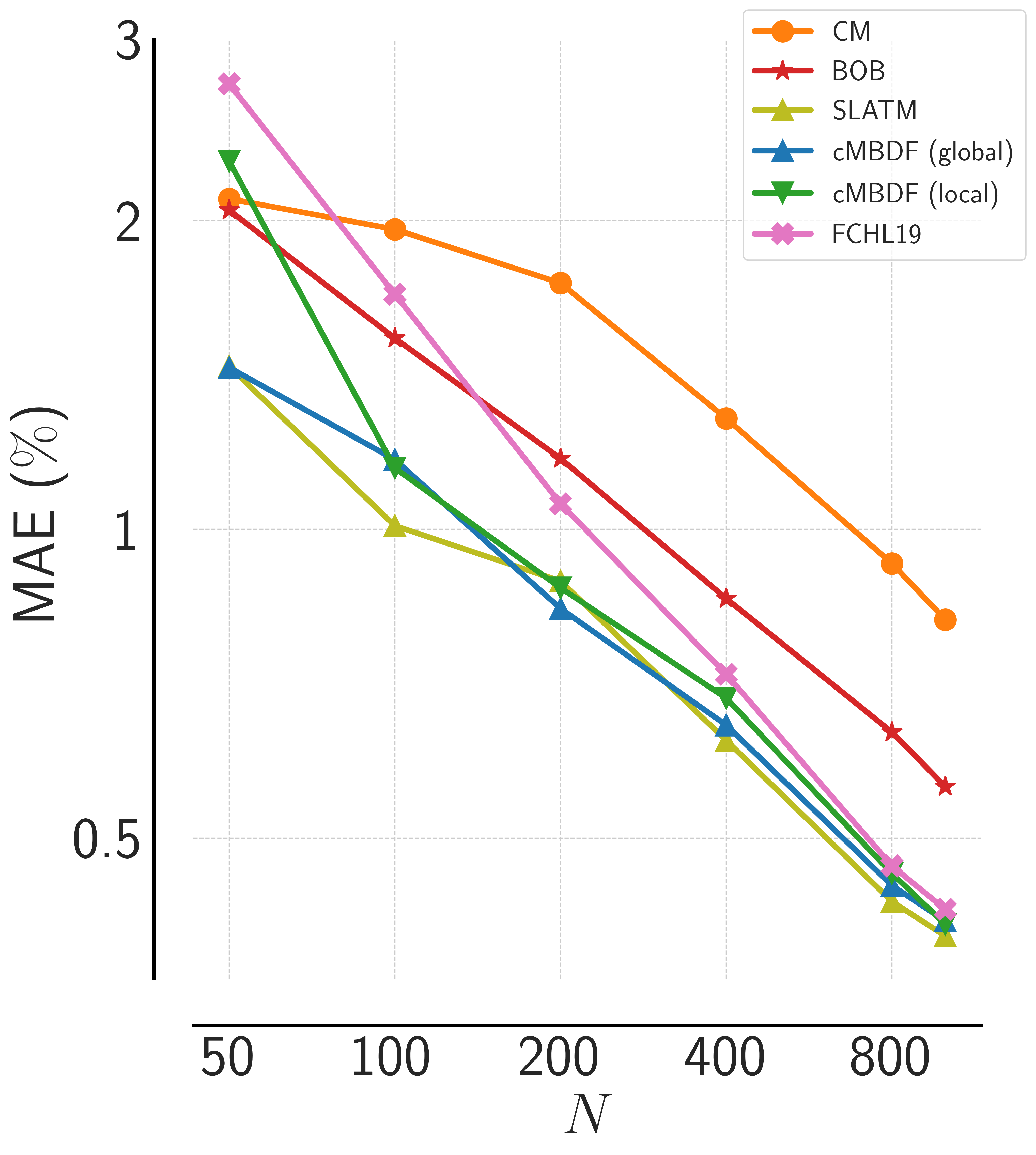}
    \caption{Learning curves showing prediction error for optimal HF admixture ratio ($a_{\mathrm{opt}}^{\mathrm{est}}$) as a function of training set size for the representations Coulomb Matrix (CM)\cite{CM_and_qm7}, Bag of Bonds (BOB)\cite{bob}, Spectrum of London and Axilrod-Teller-Muto potentials (SLATM)\cite{amons}, Faber-Christensen-Huang-Lilienfeld 19 (FCHL19)\cite{fchl19} and convolutional Many Body Distribution Functionals (cMBDF)\cite{mbdf}.
    Training and testing (200 out-of-sample amons) is performed on the set of 1169 amons from Ref.~\cite{bing_DMC}.
    It should be noted that training and prediction of the $a_{\mathrm{opt}}^{\mathrm{est}}$ label is surprisingly easier than observables like energies with simpler methods such as CM also showing a very small offset.
    }
    \label{fig:learning_curves_amons}
\end{figure}

\begin{figure}
    \centering
    \includegraphics[width=\columnwidth]{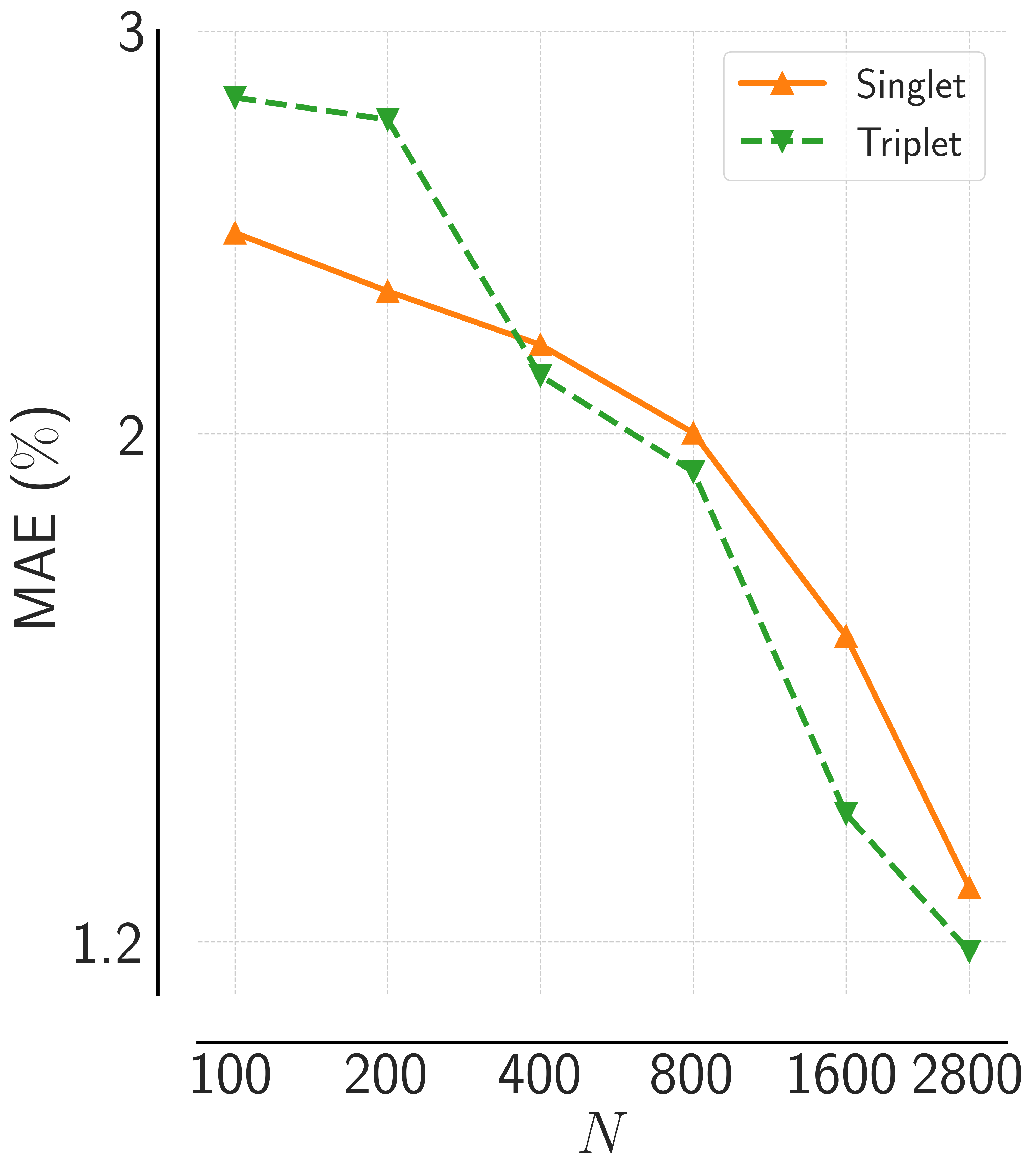}
    \caption{Learning curves showing prediction error for optimal HF admixture ratio ($a_{\mathrm{opt}}^{\mathrm{est}}$) as a function of training set size for carbenes in singlet and triplet states from the QMspin\cite{qmspin} dataset using the convolutional Many Body Distribution Functionals (cMBDF)\cite{mbdf} representation.
    }
    \label{fig:learning_curves_carbenes}
\end{figure}

\begin{table}[!htb]
\begin{center}
\begin{tabular}{ |c|c|c| } 
 \hline
 \textbf{Level of Theory} &
 \multicolumn{1}{|p{3cm}|}{\centering \textbf{MAE HOMO [eV]}} & 
 \multicolumn{1}{|p{3cm}|}{\centering \textbf{MAE LUMO [eV]}}\\ 
 \hline
 M06-2X & 1.033 & 0.726 \\
 aPBE0 (ML) & 1.181 & 0.764 \\
 $\omega$B97XD & 0.522 & 1.491 \\
 B3LYP & 2.715 & 0.607 \\
 PBE0 & 2.342 & 1.182\\
 $\mathrm{r}^{2}$SCAN & 3.474 & 0.958\\
 PBE & 3.738 & 1.267\\
 BLYP & 3.856 & 1.216\\
 \hline
\end{tabular}
\caption{HOMO and LUMO eigenvalue errors (eV) compared to GW eigen-
values for 100 molecules from the QM7b dataset~\cite{qm7b}. 
aPBE0 (ML) uses the HF exchange fraction predicted by our ML model trained on the 1169 QM9 amons from Ref~\cite{bing_DMC}}
\label{homo_lumo}
\end{center}
\end{table}

\begin{figure*}
    \centering
    \includegraphics[width=\linewidth]{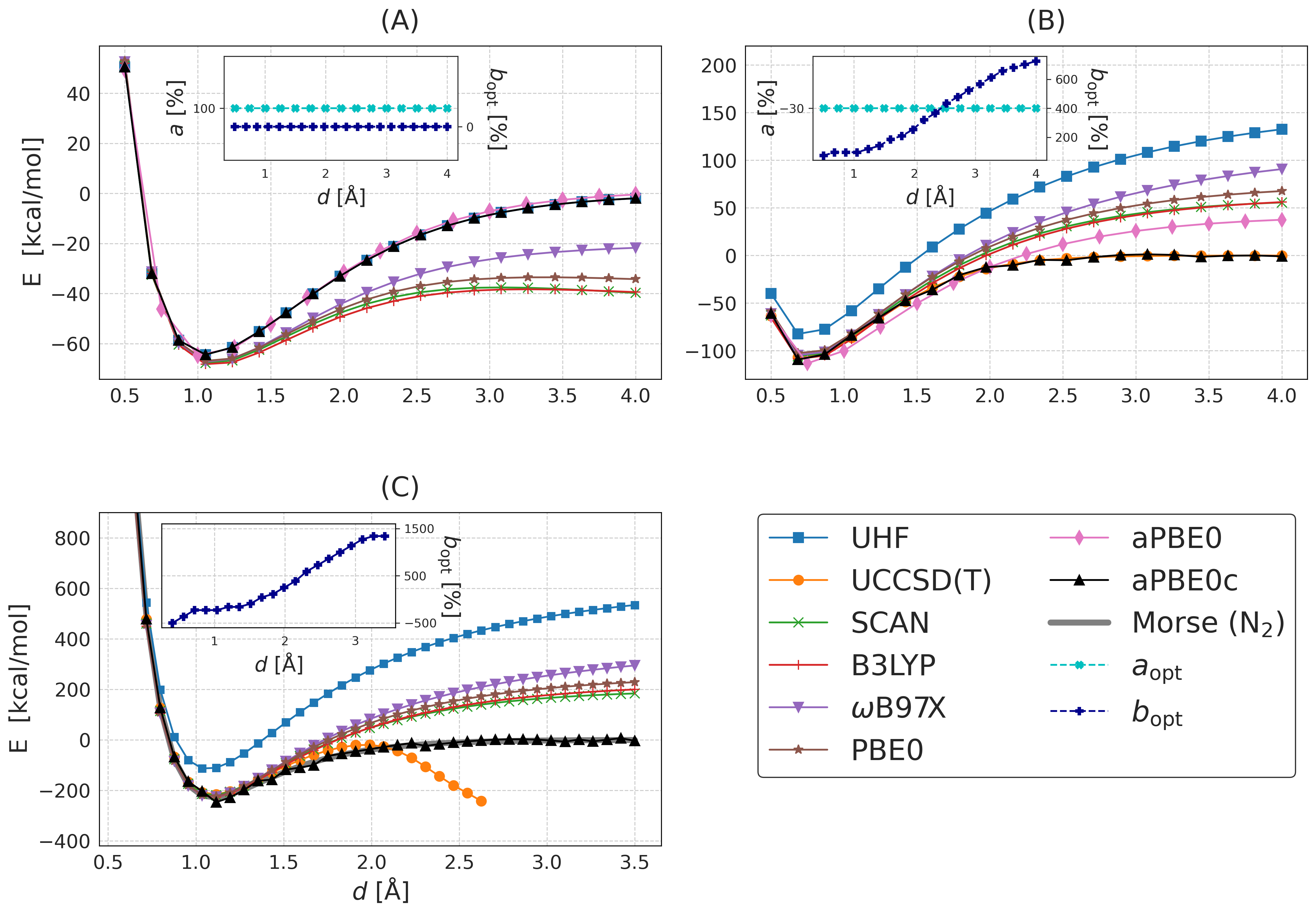}
    \caption{Dissociation energies of (A) H$_2^+$, (B) H$_2$, and (C) N$_2$ for various density functional approximations using the cc-pVTZ basis set.\cite{braida2011quantum} 
    Insets show $a_\mathrm{opt}$ and $b_\mathrm{opt}$, which correspond to optimal admixture of exact exchange or PBE correlation in the aPBE0 and aPBE0c functionals respectively. 
    Reference curves for the three cases are provided by UHF, UCCSD(T) and a Morse potential fitted to QMC data from ref.\cite{morse_N2_alastair} for H$_2^+$, H$_2$ and N$_2$ respectively.
    }
    \label{fig:all_disso}
\end{figure*}

\end{document}